\documentclass[aps,prb,twocolumn,prb,superscriptaddress]{revtex4-1}
\usepackage{graphicx}
\usepackage{appendix}

\usepackage[english]{babel} \usepackage{indentfirst}
\usepackage{amssymb,amsfonts,amsmath,bm,bbm} 
\usepackage{color}
\RequirePackage[
   hyperindex,colorlinks,bookmarksnumbered,
   plainpages=true,pdfstartview=FitH]{hyperref}
\hypersetup{linkcolor=blue,urlcolor=blue,citecolor=blue}
\usepackage{hyperref}
\usepackage{ulem}%nur fr Befehl \sout = durchstreichen
\renewcommand{\emph}[1]{\textit{#1}}%notwendig, falls Packet "ulem" verwendet wird

\newcommand{\wrt}{w.\,r.\,t.}
\newcommand{\ie}{i.\,e.}
\newcommand{\eg}{e.\,g.}

\newcommand{\qqph}{\qquad\phantom{.}}

\newcommand{\figArtab}{Fig.~\ref{fig:sample}(a),(b)}

\newcommand{\tVcr}{\widetilde V_{\rm c}^{\rm r}}

\newcommand{\tE}{\widetilde E}
\newcommand{\tVcflip}{\widetilde V_{\rm c}^{\rm flip}}
\newcommand{\Ref}[1]{Ref.~\onlinecite{#1}}

\newcommand{\Tk}{\mbox{$T_{\rm K}$}}

\newcommand{\SIAM}{{\rm SIAM}}
\newcommand{\tB}{{\widetilde B}}

\newcommand{\chiexc}{\chi_{\rm exc}^{}}

\newcommand{\Bext}{\mbox{$B$}}

\newcommand{\Omegax}{\Omega_x}
\newcommand{\chisj}{\chi_{j}}
\newcommand{\sevenKS}{.7KS}

\newcommand{\omegajmin}{\omega^{\rm min}_j}

\newcommand{\Vg}{\mbox{$V_{\rm c}$}}

\newcommand{\Vc}{\mbox{$V_{\rm c}$}}
\newcommand{\Vcpinch}{\mbox{$V_{\rm c}^{\rm pinch}$}}

\newcommand{\Vt}{\mbox{$V_{\rm t}$}}

\newcommand{\dostateszero}{{\cal A}^0}
\newcommand{\Vs}{\mbox{$V_{\rm s}$}}

\newcommand{\Ej}{E_j}
\newcommand{\Ejzero}{E_{j=0}}
\newcommand{\tVc}{\widetilde V_{\rm c}}
\newcommand{\tVcref}{\widetilde V_{\rm c}^{\rm ref}}

\newcommand{\eFermi}{\varepsilon_{\rm F}}

\newcommand{\Vcpfive}{V_{\rm 0.5}}
\newcommand{\VcQD}{V_{\rm QD}}
\newcommand{\tVczero}{\mbox{$\widetilde V_{{\rm c}0}$}}

\newcommand{\tVs}{\mbox{$\widetilde V_{\rm s}$}}

\newcommand{\Vsd}{\mbox{$V_{\rm sd}$}}
\newcommand{\Tstar}{\mbox{$T_{\ast}$}}

\newcommand{\tBstarmin}{{\widetilde B^{\rm min}_{\ast}}}
\newcommand{\Bstar}{B_{\ast}}

\newcommand{\tBstar}{{\widetilde B_{\ast}}}

\newcommand{\GQ}{\mbox{$G_{\rm Q}$}}

\newcommand{\QD}{{\rm QD}}
\newcommand{\KQD}{{\rm KQD}}
\newcommand{\QPC}{{\rm QPC}}
\newcommand{\kB}{\mbox{$k_{\rm B}$}}

\newcommand{\gel}{\mbox{$g_{\rm el}$}}
\newcommand{\muBohr}{\mbox{$\mu_{\rm B}$}}

\newcommand{\CCR}{{\rm CCR}}
\newcommand{\Ajzo}{{\cal A}_j^0(\omega)}
\newcommand{\Ajzero}{{\cal A}_j^0(0)}
\newcommand{\Tzomega}{\mathcal{T}^0(\omega)}

\newcommand{\Eq}[1]{Eq.~(\ref{#1})}

\newcommand{\Sec}[1]{Sec.~\ref{#1}}

\newcommand{\Fig}[1]{Fig.~\ref{#1}}

\newcommand{\Subfig}[2]{Fig.~\ref{#1}({#2})}

\newcommand{\Subfigsand}[3]{Figs.~\ref{#1}({#2})~and~\ref{#1}({#3})}
\newcommand{\SubFigsand}[3]{Figures~\ref{#1}({#2})~and~\ref{#1}({#3})}
\newcommand{\Subfigskomma}[3]{Figs.~\ref{#1}({#2},{#3})}
\newcommand{\Subfigsto}[3]{Figs.~\ref{#1}({#2}-{#3})}

\newcommand{\bauer}{Ref.~\onlinecite{Bauer2013}}

\newcommand{\be}{\begin{equation}}
\newcommand{\ee}{\end{equation}}
\newcommand{\beq}{\begin{eqnarray}}
\newcommand{\eeq}{\end{eqnarray}}
\newcommand{\ba}{\begin{array}}
\newcommand{\ea}{\end{array}}
\newcommand{\bea}{\begin{eqnarray}}
\newcommand{\eea}{\end{eqnarray}}

\definecolor{darkgreen}{rgb}{0,0.5,0}
\definecolor{purple}{rgb}{0.35,0,0.35}
\definecolor{orange}{rgb}{1,0.5,0}
\definecolor{darkred}{rgb}{.7,0,0}
\definecolor{darkblue}{rgb}{0,0,.3}
\definecolor{grey}{rgb}{.6,.6,.6}
\definecolor{dimgreen}{rgb}{0.2,0.6,0.1}

\newcommand{\ndSIAM}{n_{\rm d}^{\rm SIAM}}
\newcommand{\mdSIAM}{m_{\rm d}^{\rm SIAM}}

\newcommand{\mexc}{m_{\rm exc}^{}}

\newcommand{\inner}{{\rm inner}}
\newcommand{\ninner}{n_{\rm inner}^{}}
\newcommand{\minner}{m_{\rm inner}^{}}
\newcommand{\minnerQPC}{m_{\rm inner}^{\rm QPC}}
\newcommand{\minnerKQD}{m_{\rm inner}^{\rm KQD}}
\newcommand{\ninnerQPC}{n_{\rm inner}^{\rm QPC}}
\newcommand{\ninnerKQD}{n_{\rm inner}^{\rm KQD}}

\begin{document}

\title{On the relation between the 0.7-anomaly and the Kondo 
effect: \\
  Geometric Crossover between a Quantum Point Contact and a Kondo
  Quantum Dot}

\author{Jan Heyder}
\affiliation{Center for NanoScience and Fakult\"at f\"ur Physik,
Ludwig-Maximilians-Universit\"at M\"unchen, Geschwister-Scholl-Platz 1, 80539
M\"unchen, Germany}
\affiliation{Arnold Sommerfeld Center for Theoretical Physics,
Ludwig-Maximilians-Universit\"at M\"unchen, Theresienstrasse 37, D-80333 M\"unchen,
Germany}
\author{Florian Bauer}
\affiliation{Center for NanoScience and Fakult\"at f\"ur Physik,
Ludwig-Maximilians-Universit\"at M\"unchen, Geschwister-Scholl-Platz 1, 80539
M\"unchen, Germany}
\affiliation{Arnold Sommerfeld Center for Theoretical Physics,
Ludwig-Maximilians-Universit\"at M\"unchen, Theresienstrasse 37, D-80333 M\"unchen,
Germany}
\author{Enrico Schubert} 
\affiliation{Center for NanoScience and Fakult\"at f\"ur Physik,
Ludwig-Maximilians-Universit\"at M\"unchen, Geschwister-Scholl-Platz 1, 80539
M\"unchen, Germany}
\author{David Borowsky}
\affiliation{Center for NanoScience and Fakult\"at f\"ur Physik,
Ludwig-Maximilians-Universit\"at M\"unchen, Geschwister-Scholl-Platz 1, 80539
M\"unchen, Germany}
%\author{Daniela Taubert}
%\affiliation{Center for NanoScience and Fakult\"at f\"ur Physik,
%Ludwig-Maximilians-Universit\"at M\"unchen, Geschwister-Scholl-Platz 1, 80539
%M\"unchen, Germany}
\author{Dieter Schuh}
\affiliation{Institut f\"ur Angewandte Physik, Universit\"at Regensburg, D-93040
Regensburg, Germany }
\author{Werner Wegscheider}
\affiliation{Laboratory for Solid State Physics, ETH Z\"urich, CH-8093 Z\"urich,
Switzerland}
\author{Jan von Delft}
\affiliation{Center for NanoScience and Fakult\"at f\"ur Physik,
Ludwig-Maximilians-Universit\"at M\"unchen, Geschwister-Scholl-Platz 1, 80539
M\"unchen, Germany}
\affiliation{Arnold Sommerfeld Center for Theoretical Physics,
Ludwig-Maximilians-Universit\"at M\"unchen, Theresienstrasse 37, D-80333 M\"unchen,
Germany}
\author{Stefan Ludwig}
\affiliation{Center for NanoScience and Fakult\"at f\"ur Physik,
Ludwig-Maximilians-Universit\"at M\"unchen, Geschwister-Scholl-Platz 1, 80539
M\"unchen, Germany}

\date{\today}
%\date{June 20, 2012}

%\baselineskip15pt

 \begin{abstract}
   Quantum point contacts (QPCs) and quantum dots (QDs), two
   elementary building blocks of semiconducting nanodevices, both
   exhibit famously anomalous conductance features: the 0.7-anomaly in
   the former case,
%\cite{Thomas1996,Thomas1998,Reilly2001,
%     Cronenwett2002,%Chung2007,
%     Koop2007,Pepper2008,Nakamura2009,Komijani2010,
%     Smith2011,Bauer2013}, 
the Kondo effect in the latter.
%\cite{%Glazman1988,Ng1988,Goldhaber-Gordon1998Nature,
%Cronenwett1998,Wiel2000,Nygard2000,%Jarillo-Herrero2005,
%Kretinin2011}.  
For both the 0.7-anomaly and the Kondo effect, the conductance shows a
remarkably similar low-energy dependence %\cite{Cronenwett2002}
on temperature $T$, source-drain voltage $\Vsd$ and magnetic field
$B$. In a recent publication [F. Bauer \textit{et al.}, Nature,
\textbf{501}, 73 (2013)], we argued that the reason for these
similarities is that both a QPC and a KQD feature \textit{spin
  fluctuations} that are induced by the sample geometry, confined in a
small spatial regime, and enhanced by interactions.
% geometry-induced, interaction-enhanced,  spatially-localized spin
%  fluctuations. 
Here we further explore this notion experimentally and theoretically
by studying the geometric crossover between a QD and a QPC, focussing
on the $B$-field dependence of the conductance. We introduce a
one-dimensional model that reproduces the essential features of the
experiments, including a smooth transition between a Kondo QD and a
QPC with 0.7-anomaly.  We find that in both cases the anomalously
strong negative magnetoconductance goes hand in hand with strongly
enhanced local spin fluctuations.  Our experimental observations
include, in addition to the Kondo effect in a QD and the 0.7-anomaly
in a QPC, Fano interference effects in a regime of coexistence between
QD and QPC physics, and Fabry-Perot-type resonances on the conductance
plateaus of a clean QPC. We argue that Fabry-Perot-type
  resonances occur generically if the electrostatic potential of the QPC
 generates a flatter-than-parabolic barrier top.
 \end{abstract}

 \maketitle

\section{Introduction}

A QPC is a narrow one-dimensional (1D) constriction and a QD a small
isolated puddle of charges, patterned in a two-dimensional electron
system (2DES), e.\,g.\ by applying voltages to local gates.  Being key
ingredients of semiconductor-based quantum circuits, much effort has
been devoted to understand their behavior at a fundamental level.
Here, we investigate the geometric crossover between a QPC and a
QD. The motivation for this study is to shed light on similarities and
differences between the 0.7-anomaly exhibited by the conductance of a
QPC, and the Kondo effect found in a Kondo quantum dot (KQD) that
hosts an odd number of electrons and hence contains a localized spin.

The linear conductance $G(\Vg)$ of a QPC is famously quantized in
units of $\GQ = 2e^2/h$, when measured as function of the gate voltage
$\Vc$ defining the channel width 
\cite{Wees1988,Wharam1988,Buettiker1990}.  The 0.7-anomaly is observed
as an additional shoulder when the dimensionless conductance,
$g=G/\GQ$, reaches the value $g \simeq 0.7$ in the first conductance
step\cite{Thomas1996,Thomas1998,Reilly2001,
  Cronenwett2002,Chung2007,Koop2007,Pepper2008,Nakamura2009,Komijani2009,
  Smith2011,Bauer2013}. It shows strikingly anomalous behavior as
function of temperature ($T$), magnetic field ($B$) and source-drain
voltage ($\Vsd$), which can not be explaqined within a non-interacting
model.  The \textit{low-energy} $T$-, $B$- and $\Vsd$-dependencies of
the 0.7-anomaly are similar to those of a KQD
\cite{Glazman1988,Ng1988,Goldhaber-Gordon1998Nature,Cronenwett1998,Wiel2000,Nygard2000,Jarillo-Herrero2005,Kretinin2011}
at excitation energies well below its Kondo temperature, $\Tk$: for
both QPC and KQD, the linear conductance strongly decreases with
increasing $B$ and $T$, while the non-linear conductance shows a
zero-bias peak as function of $\Vsd$, that splits into two subpeaks
with increasing $B$.  We will call this similar behavior the
``0.7-Kondo-similarity'' (\sevenKS). 
%\comment{Every resonance (not
%  only Kondo, but even in the non-interacting case) shows all three
%  above mentioned features (in fact for $G=\GQ$ - which holds for the
%  Kondo Effect and, according to Meir, also in the $T=B=0$ limit for
%  the 0.7 anomaly - the conductance can only drop as a function of
%  external parameters). The same holds for the QPC: We show in nature,
%  that all features (+ the exponential dependence of $\Bstar$ and the
%  enhanced susceptibility in a similar $V_c$-regime) are present even
%  without interactions. Thus, the whole \sevenKS~ can be explained at
%  $U=0$ in the QPC. Contrary, an explanation for 0.7 anomaly does need
%  them. To me this should be made clear. Otherwise we weaken our own
%  work drastically and give in to the Kondo scenario of Meir too
%  much. } 
To explain it, Meir and collaborators
\cite{Meir2002,Golub2006,Rejec2006} have argued that a ``quasi-bound
state'' in the QPC, predicted via spin-density-functional theory,
harbors a localized spin that causes Kondo-like conductance anomalies.

We have recently proposed a scenario that explains the microscopic
origin of the 0.7-anomaly and of the \sevenKS\  without invoking
a localized spin \cite{Bauer2013}.  In a nutshell, we argue
that the 0.7-anomaly is a direct consequence of a ``van Hove ridge'',
\ie\ a smeared van Hove peak in the local density of states (LDOS) at
the bottom of the lowest 1D sub-band of the QPC, whose shape follows
that of the QPC potential barrier.  Invoking a semi-classical picture,
the LDOS is inversely proportional to the velocity of an electron with
given energy at a given position; the van Hove ridge, which
corresponds to a locally enhanced LDOS, thus reflects the fact that
electrons are being slowed down while they cross the 1D barrier
constituting the QPC. The slow electrons experience strongly enhanced
mutual interactions. When the QPC barrier is tuned to lie just below
the chemical potential, transport properties are significantly
affected by these strongly enhanced electron interactions.  In \bauer\
we have shown that this amplification of interaction effects is
sufficient to fully explain the 0.7-anomaly.

The above-mentioned two scenarios for explaining the 0.7-anomaly,
evoking a quasi-localized state or a van Hove ridge, respectively,
have an important common feature, namely that in both cases, the
physics is governed by slow electrons above the barrier top.  In this
regard, it is not surprising that both scenarios are compatible in
their predictions for the low-energy behavior of the 0.7-anomaly,
which, as mentioned above, is similar to that of the Kondo effect. In
\bauer, we attributed this \sevenKS\ to the fact that both a KQD and a
QPC involve a spin-singlet ground state featuring \textit{spatially
  confined spin fluctuations.}  While for a KQD they are associated
with the screening of a truly localized spin, for a QPC they result
from the extended but curved structure of the van Hove ridge and
include a large number of spins.
%;
%  but this distinction, which is important on short length scales
%  (high energies), does not matter on long ones (low energies). 
  In both cases, these spin fluctuations are characterized by an
  exponentially small energy scale, called $\Bstar$ in \bauer, which
  goes hand in hand with an enhanced local spin susceptibility.  For a
  KQD this low-energy scale corresponds to the Kondo temperature,
  $\Tk$.
%\comment{not sure i follow:
%  Above we say, that our scenario yields a similar low-energy behavior
%  as Kondo physics. So, if we now state a different $V_c$-dependence
%  we should definitely explain why: "
%For a KQD the quadratic
%  exponential dependence of the scale $\Tk$ is interaction-induced,
%  while for a QPC with smooth parabolic effective potential the linear
%  exponential scale $\Bstar^\QPC$ is defined through geometry and only
%  modulated by interactions"}
The \sevenKS\ pertains to energies well
below $\Bstar$; we have argued in \bauer\ that it results from the
fact that for such low energies, both a KQD and a QPC show
Fermi-liquid behavior of the type associated with quasi-particles
experiencing spatially localized interactions.  For the Kondo effect,
a corresponding Fermi-liquid theory has been worked out by
Nozi\`eres\cite{Nozieres1974}; doing the same for a QPC would be an
interesting challenge for the future.

Though slow electrons form the common ground for both the van Hove
ridge scenario and quasi-localized state scenario, the two scenarios
differ substantially in their microscopic description of the slow
electrons' dynamics. The van Hove ridge scenario describes them via
the LDOS, thus incorporating the geometric shape of the barrier. In
contrast, the quasi-localized state scenario describes them more
simplistically in terms of a magnetic moment, i.e.\ a truly localized
state, thus arriving at a seemingly simpler model, akin to the
single-impurity Anderson model. This apparent simplification, however,
comes at a price: the physics of the Anderson model involves a free
local moment high energies, and Fermi-liquid behavior emerges only at
low energies, when the local moment is screened.  For a QPC such a
``detour'' (first evoke a local moment, then argue that it is
screened) is in our opinion not needed: in our van Hove ridge
scenario, Fermi-liquid behavior is present \textit{a priori}.
Moreover, in \bauer\ we have found no indications that a smooth
parabolic barrier hosts a discrete, truly localized spin, and no
similarities (in our experimental results or theoretical predictions)
between the Kondo effect and the 0.7-anomaly at high energies
($\gtrsim\Bstar$), where the Kondo effect is governed by an unscreened
local moment. This shows that when the ``slow electrons'' in a
  QPC are probed at energies $\gtrsim\Bstar$, they do behave
  differently from the magnetic moment in a KQD. (In
  Sec.~\ref{sec:magnetization} we offer additional evidence for this
  conclusion by comparing the behavior of the magnetization of a KQD
  and a QPC at large magnetic fields.)
  
  The differences between a KQD and a QPC come to the fore very
  explicitly in the functional dependence of the low-energy scale
  $\Bstar$ on system parameters such as the gate voltage and the
  interaction strength (discussed in detail in Sec.~\ref{sec:Bstar}
  below).
%For both a KQD and a QPC, $\Bstar$ arises
%through an interplay of geometry and interactions; however, since the
%detailed mechanisms leading to $\Bstar^\KQD$ and $\Bstar^\QPC$ are
%different, their functional dependences on, for example, gate voltage,
%are also different: 
  For a KQD, the scale $\Bstar^\KQD$ can not meaningfully be defined
  in the absence of interactions (since then no local moment forms),
  and $\ln \Bstar^\KQD$ depends quadratically on gate
  voltage\cite{Haldane1978,Wiel2000,Kretinin2011}. For a QPC, in
  contrast, $\Bstar^\QPC$ can be meaningfully defined even in the
  absence of interactions, and $\ln \Bstar^\QPC$ depends linearly on
  gate voltage.  When interactions are turned on, $\Bstar^\QPC$ is
  reduced strongly, but its functional dependence on gate voltage
  hardly changes (see \bauer, Sec.~S-5).

  The present paper aims to elaborate the relation between local spin
  fluctuations and the \sevenKS\ in more detail, and, more generally,
  to analyse the similarities and differences between the Kondo effect
  and the 0.7-anomaly, focussing on their dependence on magnetic field
  at low temperature, in equilibrium.  We experimentally and
  theoretically study the smooth \textit{geometric crossover} between
  a \KQD\ and a QPC, and hence between the Kondo effect and the
  0.7-anomaly.  Experimentally, we measure the conductance throughout
  the QD-QPC crossover using a highly tunable nanostructure
  tailor-made for this purpose.  In our theoretical work, we consider
  a 1D model with local interactions and a smooth potential barrier,
  similar to that used in \bauer, but now tune the shape of the
  potential barrier in such a way that it smoothly crosses over
  between a single barrier, representing a QPC, and a double barrier,
  representing a KQD. We use the functional renormalization group
  (fRG)\cite{Andergassen2006,Karrasch2006a,Metzner2012,Bauer2014} to
  calculate how transport and thermodynamic properties at $T=0$ change
  during this crossover.  This allows us to track the extent to which
  features characteristic for Kondo correlations do or do not survive
  in the QPC regime. A central finding is that the strongly enhanced
  local spin susceptibility in the center of the system that is found
  for both a QPC and a KQD goes hand in hand with an anomalously
  strong magnetic field dependence of the conductance. This is
  actually not surprising, since a large spin susceptibility indicates
  a strong depletion of that spin species that is energetically
  disfavoured in the presence of a small magnetic field. Our analysis
  pinpoints the enhanced local spin susceptibility as the common
  feature of both systems that underlies the \sevenKS\ regarding its
  dependence on magnetic field.

  The paper is organized as follows: \Sec{sec:setup} describes our
  experimental setup and our measurements for the QD-QPC
  crossover. \Sec{sec:model} presents the model by which we describe
  this crossover, discusses how the geometry of the QPC or QD barrier
  influences the noninteracting LDOS and noninteracting transmission
  probability, and summarizes the key elements of our fRG approach for
  treating interactions. \Sec{sec:magnetoconductance} compares fRG
  results and experimental data for this crossover, showing that our
  model captures its main features in a qualitatively correct manner.
  \Sec{sec:localproperties} presents the results of fRG calculations
  for local properties, such as the local density, magnetization and
  spin susceptibility for both a QPC and a KQD, and for the spin
  susceptibility during the QPC-QD crossover, which very clearly
  reveals the origin of the \sevenKS. \Sec{sec:magnetization} presents
  fRG results on the evolution of the magnetization with $\tB$,
  highlighting the difference between a KQD and QPC when probed at
  energies beyond $\Bstar$.  \Sec{sec:conclusions} offers a summary
  and outlook.  An appendix presents and discusses a movie with fRG
  results that show how the conductance evolves with magnetic field
  during the QD-QPC crossover.

% \papertwo{In a follow-up paper\cite{Delft2014}, we will continue our
%  comparative study of the Kondo effect and the 0.7-anomaly by
%  presenting additional theoretical results, obtained by fRG.  There
%  we will discuss the similarities and differences between the
%  transmission phase shifts of a QPC and a KQD, calculated by fRG; we
%  will elucidate to what extent a future Fermi-liquid theory for the
%  0.7-anomaly would be similar to or different from that known for the
%  Kondo effect.}

 \section{Experimental Setup}
\label{sec:setup}

\begin{figure}[ht]
\centering
  \includegraphics[width=\linewidth]{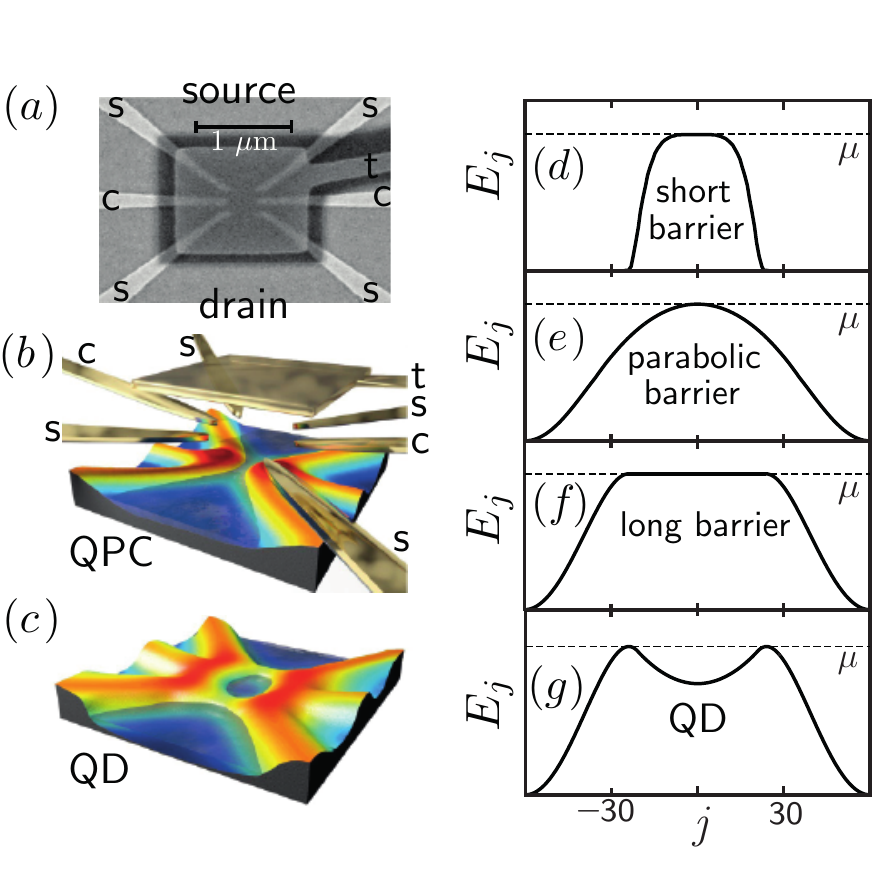}
  \caption{Geometric crossover between QPC and QD -- sample and shape
    of effective potential. (a) Scanning electron microscope picture
    of the gate layout, which features a top gate at voltage $\Vt$,
    two central gates at voltage $\Vc$, and four side gates at voltage
    $\Vs$. Negative voltages $ \Vc$ and $\Vs$ deplete the 2DES 85\,nm
    beneath the sample surface, inducing a tunable effective
    electrostatical potential landscape there. (b,c) Artist's
    depiction of this landscape for a QPC and QD, respectively
    [red/yellow:\ high electrostatic potential; blue:\ low potential,
    Fermi sea darkened; golden structures at top of (b): gates]. (d-g)
    The effective potential $E_j$ of the Hamiltonian given in
    \Eq{eq:model}, shown for four different choices of the barrier 
shape (black lines), tuned by adjusting the 
central
 barrier height $\tVc = \Ejzero - \eFermi$, the side barrier height
 $\tVs$ (with $\widetilde V_{\rm c,s} \propto - |e| V_{\rm
     c,s}$), and the barrier width. 
These choices correspond to 
(d) a short QPC with a flat potential top ($\tVc = \tVs$); (e) a QPC
described by a parabolic potential top ($\tVc > \tVs$); (f) a long QPC
with a flat potential top ($\tVc = \tVs$); and (f) a QD $(\tVc <
\tVs$).  
\label{fig:sample}}
\end{figure}
We use the multigate layout shown in \Subfig{fig:sample}{a} to
laterally define a nanostructure in the two-dimensional electron
system (2DES) located 85\,nm beneath the surface of our GaAs/AlGaAs
heterostructure. The low temperature charge carrier density is $1.9
\times 10^{11}\,\mathrm{cm}^{-2}$ and the mobility $1.2 \times
10^{6}\,\mathrm{cm}^{2}/\mathrm{Vs}$. Magnetic fields are aligned
parallel to the 2DES and to the 1D channel defining the QPC (current
direction).  The field's alignment is optimized by use of a two-axis
magnet and controlled by magnetotransport measurements. The electron
temperature in all measurements presented here is $T_0\simeq30\,$mK
according to our estimations from separate temperature dependent
measurements (not shown, see also \bauer).

Seven gates provide a particularly high tunability of the central
constriction region (CCR) of our device, located at the center of
\Subfig{fig:sample}a between the tips of six gates. We apply one
voltage, say \Vc, to both central gates, and another, say \Vs, to all
four side gates.  Our sample also contains a global top gate [see
\figArtab], electrically insulated from other gates by a layer of
cross-linked PMMA (plexiglass). The top gate can be used to adjust the
carrier density of the 2DES in the contacts of the CCR and thereby
control the effective interaction strength between electrons
\cite{Bauer2013}.  In this article we keep $\Vt$ fixed at 0.8\,V.  By
suitably tuning $\Vc$ and \Vs, we are able to smoothly reshape the
potential landscape in the 2DES in such a way that it crosses over
from a saddle point potential defining a QPC [\Subfig{fig:sample}{b}]
to a symmetric local minimum defining a QD
[\Subfig{fig:sample}{c}]. The corresponding effective 1D potential
barrier shape crosses over from a single barrier
[\Subfigsto{fig:sample}{d}{f}], whose top is parabolic only in a
relative narrow range of gate voltages, to a symmetric double barrier
[\Subfig{fig:sample}{g}].

\begin{figure*}[]
\includegraphics[width=\linewidth]{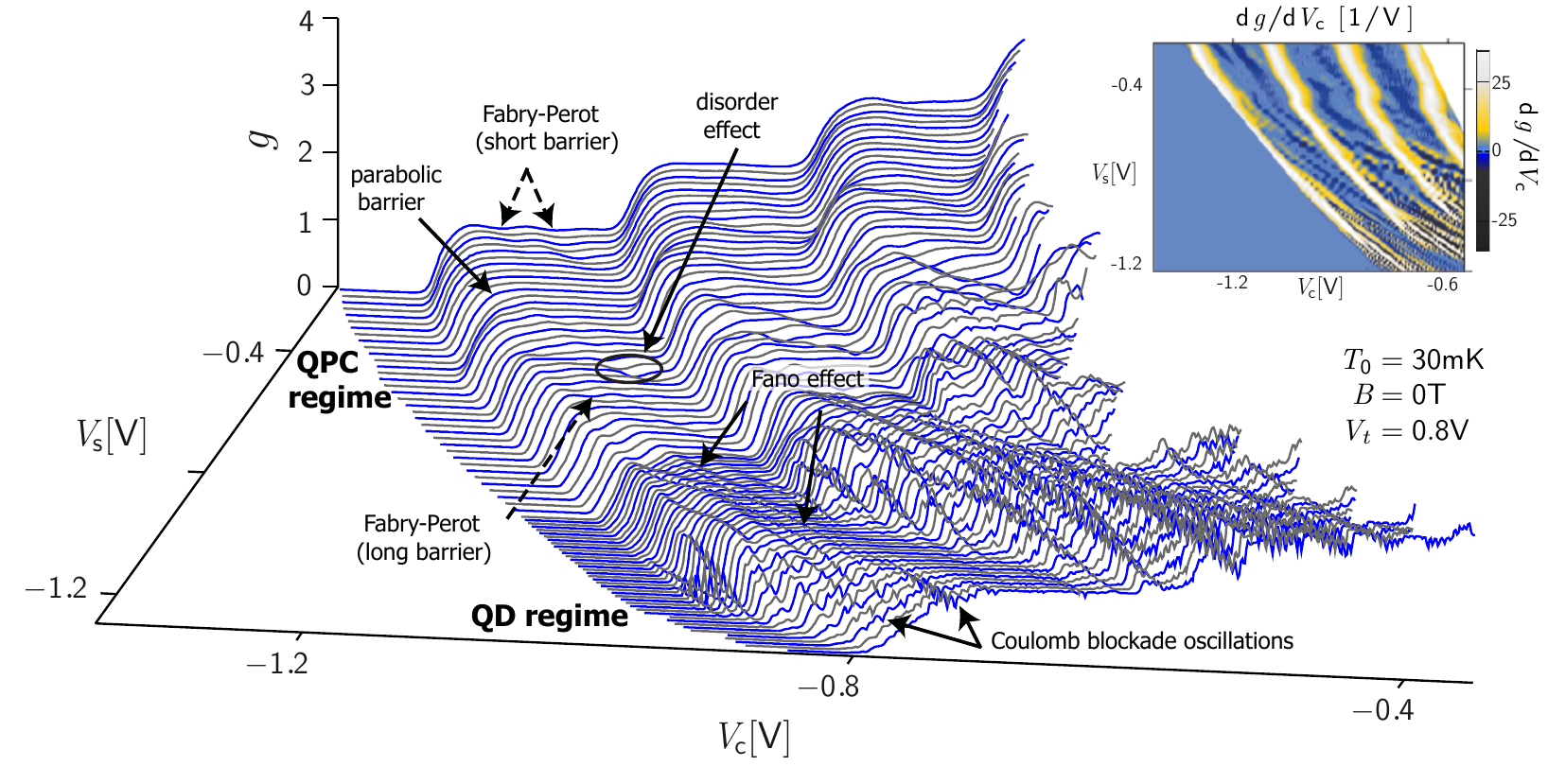}
\caption{The linear-response differential conductance $g$ (main panel)
  and transconductance $dg/d\Vc$ (inset) as a function of $\Vc$ and
  $\Vs$, showing the geometric crossover between a QD and a QPC.
  Detailed measurements of the 0.7-anomaly at fixed side gate voltage,
  reported in \bauer\ and shown in part in \Subfig{fig:magnetocond}{d}
  below, were performed at $\Vs=-0.4\,$V, where the first plateau is
  absolutely flat, implying a parabolic barrier top. Features marked
  by arrows are explained in the main text.
\label{fig:rawdata}}
\end{figure*}

Experimentally we have studied this smooth transition from a QD to a
QPC by measuring the two-terminal differential conductance $g$ via the
linear response of the current to small modulations of the applied
source-drain voltage.  \Fig{fig:rawdata}
shows raw data (albeit already corrected for the lead resistances)
measured at $B=0$ as a function of $\Vc$ and $\Vs$.  For our most
negative $\Vs$-values and near pinch-off, \ie\ also negative $\Vc$,
the conductance traces show pronounced Coulomb blockade oscillations
as function of $\Vc$ (at the bottom center part of the
figure).  This indicates that the CCR constitutes a single,
well-defined QD with a substantial Coulomb charging energy.  When
$\Vs$ is made less negative the Coulomb blockade oscillations
disappear altogether. The reason is that the local electrostatic
potential near the side gates decreases and eventually becomes smaller
than the electrostatic potential between the center gates,
corresponding to a transition from a double barrier potential as in
\Subfig{fig:sample}{g} to a single barrier top as in
\Subfigsto{fig:sample}{d}{f}. In the process the QD disappears, and
with it the localized states, and a clean QPC remains. Its barrier top
may or may not be parabolic, depending on the value of $\Vs$.  At
$\Vs\simeq-0.4\,$V, $g(\Vc)$ clearly shows several smooth conductance
steps as function of $\Vc$, as expected for the pinch-off curves of a
clean parabolic QPC.

The broad transition regime between QPC and QD displays a combination
of both 1D conductance steps and Coulomb blockade oscillations. The
latter are most pronounced at the steps between conductance plateaus
($0<g<1 $, $1 <g<2 $, \dots) and occur in clusters with a rather
similar structure, as can be best seen in the inset of
\Fig{fig:rawdata}, which shows $\mathrm d g / \mathrm d\Vc$ using a
color scale. This repeating pattern of Coulomb blockade oscillations
indicates a coexistence of a QD in the not yet (fully) occupied
one-dimensional subband of the CCR with already fully occupied lower
one-dimensional subbands contributing to QPC-behavior. This causes the
charge configurations of the QD to repeat at adjacent QPC conductance
steps when the number of occupied one-dimensional subbands changes by
one. Note that as $\Vs$ becomes more negative, the spacing between
Coulomb blockade peaks within each cluster tends to increase (causing
the cluster to ``fan out'', see \Fig{fig:rawdata}, inset, bottom right
corner).  This reflects an increase in the QD charging energy, brought
about by the steepening of the confinement potential when its side
barriers become higher.

Whenever a Coulomb blockade oscillation enters a conductance plateau
at $g=1,2,\dots$ the corresponding conductance maxima of the QD turn
into narrow conductance dips, some of which are marked by solid arrows
in \Fig{fig:rawdata}.  We interpret these dips as Fano resonances
between the 1D channel of the QPC and localized states of the QD.

We also observe broader and very shallow conductance oscillations on
the conductance plateaus deeper in the QPC regime (at larger $\Vs$),
marked by dashed arrows in \Fig{fig:rawdata}. They are absent only in
a narrow region around $\Vs\simeq-0.4\,$V (on the first plateau), and
the oscillation period observed for $\Vs < -0.4\,$V is shorter than
that observed for $\Vs>-0.4\,$V.  We interpret these as
Fabry-Perot-like resonances that arise whenever the barrier shape is
not parabolic: as $\Vs$ is increased within the regime of a clean QPC,
starting from around $\Vs \simeq - 0.8\,$V, we observe a transition
from a long flat barrier via a parabolic barrier near
$\Vs\simeq-0.4\,$V to a short flat barrier for $\Vs>-0.4\,$V,
  where also $\Vs\gg\Vc$, (as sketched in
\Subfigsto{fig:sample}{d}{f} and indicated in \Fig{fig:rawdata}). We
will discuss the origin and behavior of these Fabry-Perot resonances
in more detail in sections \ref{subsec:LDOS+Transmission} and
\ref{sec:fabry-perot} below.

At a particular side gate voltage, near $\Vs \simeq -0.6$\,V, two
conductance traces show a marked dip (indicated by an ellipse) near
the end of the first plateau.  We interpret this distinct reduction of
the conduction as reflection of electrons caused by disorder in
  the form of  a distinct defect.

  The multi-gate tunability of our device has the important advantage
  that it allows these type of effects (Fabry-Perot and/or disorder)
  to be avoided, if desired.  Indeed, at side gate voltages near $\Vs
  \simeq -0.4$\,V, no such effects are seen around the first
  conductance plateau is completely flat. We have therefore used $\Vs
  = -0.4$\,V for the detailed measurements of the 0.7-anomaly reported
  in \bauer, some of which are also shown in
  \Subfig{fig:magnetocond}{d} below. An additional option would be to
  apply various different voltages to the individual four side gates
  or the two central gates to overcome possible disorder
  effects. However, the high quality of our sample rendered such
  options unnecessary, allowing us to maintain a high degree of
  symmetry of the electrostatic potential defining the CCR.

\Fig{fig:rawdata} displays two additional remarkable trends: (i) the
quantized plateaus in $g(\Vc)$ become wider as $\Vs$ is
increased. (ii) This goes along with an increase of the step width
between plateaus as is best seen in the inset of \Fig{fig:rawdata}
(consider the width of the white-yellow bands in the upper half of the
plot, the QPC regime). Trend (i) indicates that the lateral
confinement becomes stronger with more positive $\Vs$, leading to a
larger characteristic energy spacing between the 1D subbands. This
also implies a larger on-site exchange energy, $U$, between the
electrons. Trend (ii) confirms our statement, above, that the width of
the barrier, seen by electrons, decreases as $\Vs$ is increased,
because a narrower barrier causes the step width to
  become wider (see the discussion in \Sec{subsec:LDOS+Transmission}
  and \Subfigsto{fig:LDOS-QPC-QD}{a}{c} below). Both trends together
(which further depend on the topgate voltage, see Supplementary
Material in \bauer) provide us with an experimental toolkit to
precisely measure the geometry dependence of the
0.7-anomaly. This could be used for a detailed test of the
  predictions of our model in \bauer. A  study of this kind is
  beyond the scope of this work and left for the future.

\section{Theoretical Model} 
\label{sec:model}

In this section we present the one-dimensional model used for our
theoretical description of the QD-QPC crossover, featuring a smooth
barrier and short-ranged interactions [\Sec{subsec:Hamiltonian}].  We
first illustrate its geometrical properties in the absence of
interactions, by showing results for the noninteracting LDOS and
noninteracting transmission [\Sec{subsec:LDOS+Transmission}]. Within
this non-interacting framework we explain why Fabry-Perot-type
resonances occur whenever the barrier top is not parabolic
[\Sec{sec:fabry-perot}]. Finally, we summarize the key ingredients of
the fRG approach used here to treat interaction effects
[\Sec{subsec:fRG}].

\subsection{Hamiltonian}
\label{subsec:Hamiltonian}

To describe the QD-QPC crossover we restrict ourself to the lowest 1D
subband of the CCR and adopt the model introduced in
  \bauer\ (see its Supplementary Information,
  Section~S-4.B, ``model I''), whose notational conventions we adopt
  here, too.  The Hamiltonian has the form
\begin{eqnarray}
H = \sum_{j\sigma} \left[
E_{j\sigma} \hat n_{j\sigma}
- \tau ( d^\dagger_{j+1 \, \sigma} d_{j \sigma} + \rm{h.c.}) \right]
+ \sum_j U_j n_{j\uparrow} n_{j \downarrow} ,
\nonumber
\\
\label{eq:model}
\end{eqnarray}
where $\hat n_{j \sigma} = d^\dagger_{j \sigma} d_{j\sigma}$ counts
the number of electrons with spin $\sigma$ ($= \! \pm$ for
$\uparrow$,$\downarrow$) at site $j$. It describes an infinite
tight-binding chain with constant lattice spacing $a = 1$ (taken as
length unit), constant hopping amplitude $\tau =1$ (taken as energy
unit), on-site interaction $U_j$ and on-site potential energy $E_{j
  \sigma}= \Ej - \frac{\sigma}{2}\tB$.  Here $\Ej = \tE(ja)$ models
the smooth electrostatic potential $\tE(x)$ defined by gates, and the
Zeeman energy $\tB$ accounts for a uniform external parallel magnetic
field.  (We use tildes to distinguish model parameters from
experimental ones, with $\tB = |\gel| \muBohr \Bext$ for the magnetic
field, where $\gel < 0$ for GaAs, $\tilde T = k_{\rm B} T$ for
temperature, and $\widetilde V_{\rm c,s} \propto - |e| V_{\rm c,s}$
for the central and side gate voltages.)  We neglect spin-orbit
interactions and other orbital effects. We take $U_j$ and $\Ej$ to be
nonzero only within a central constriction region (\CCR) of $N=2N' +1
$ sites around $j=0$, representing the QD or QPC.  The rest of the
chain represents two noninteracting leads with effective mass $m =
\hbar^2/(2 \tau a^2)$ (defined as the curvature of the dispersion at
the band bottom in the bulk), chemical potential $\mu$ and bulk Fermi
energy $\eFermi = 2 \tau + \mu$; we choose $\mu=0$, implying
half-filled leads.  $U_j$ is set to a constant value $U$ within the
CCR for all but the outermost sites of the \CCR, where it drops
smoothly to zero.  (For an explicit
formula for $U_j$, see Eq.~(S14) of the Supplementary Information of
\bauer.)  The shape of $\Ej$ is governed by two parameters, $\tVc $
and $\tVs$, that respectively mimic the effects of the central and
side gates in experiment.  $\tVc < \tVs$ defines a QD with side
barrier height $\tVs$ \wrt\ $\mu$ [\Subfig{fig:sample}{f}].  $\tVc >
\tVs$ yields a QPC with a single central barrier
[\Subfig{fig:sample}{d}].  Its shape near the top is chosen to be
parabolic \cite{Buettiker1990}, unless stated otherwise
[\Fig{fig:nonparabolic-barrier} features non-parabolic barrier
tops]. We parametrize parabolic barrier tops by
  \begin{eqnarray}
    \label{eq:QPCbarriertop}
    \tE(x) \simeq \tVc + \eFermi - %\frac{\Omegax^2}{4 \tau} j^2 \; 
 \frac{m \Omegax^2 x^2}{2 \hbar^2} \; . 
  \end{eqnarray}
  Here $\tVc$ is the barrier height measured \wrt\ the chemical
  potential, and the barrier curvature is characterized by an energy
  scale $\Omegax$. We emphasize that by ``parabolic'' barrier,
    we mean that the quadratic $x$-dependence of \Eq{eq:QPCbarriertop}
    holds over an energy range of at least $\Omegax$ from the barrier
    top (i.e.\ up to $x$-values large enough that $E(0)- E(x) \gtrsim
    \Omegax$).  Then the width of the conductance step is given by
    $\Omegax$. 

Typical choices of the potentials described by $E_j$ are shown in
\Subfigsto{fig:sample}{d}{g}, \Subfigsto{fig:LDOS-QPC-QD}{a}{f} and to
some extent \Fig{fig:nonparabolic-barrier} (inset). For situations
where the shape of $E_j$ is not shown explicitly
[\Subfigsto{fig:QD-QPC}{a}{c}, \Fig{fig:magnetocond},
\Subfigsto{fig:local}{i}{l}], it is chosen according to a formula
given by Eq.~(S15) of the Supplementary Information of \bauer, with
$j_s = 60$ and $N'= 150$ there.

\subsection{Noninteracting LDOS and transmission}
\label{subsec:LDOS+Transmission}

\begin{figure*}[t]
  \includegraphics[width=\linewidth]{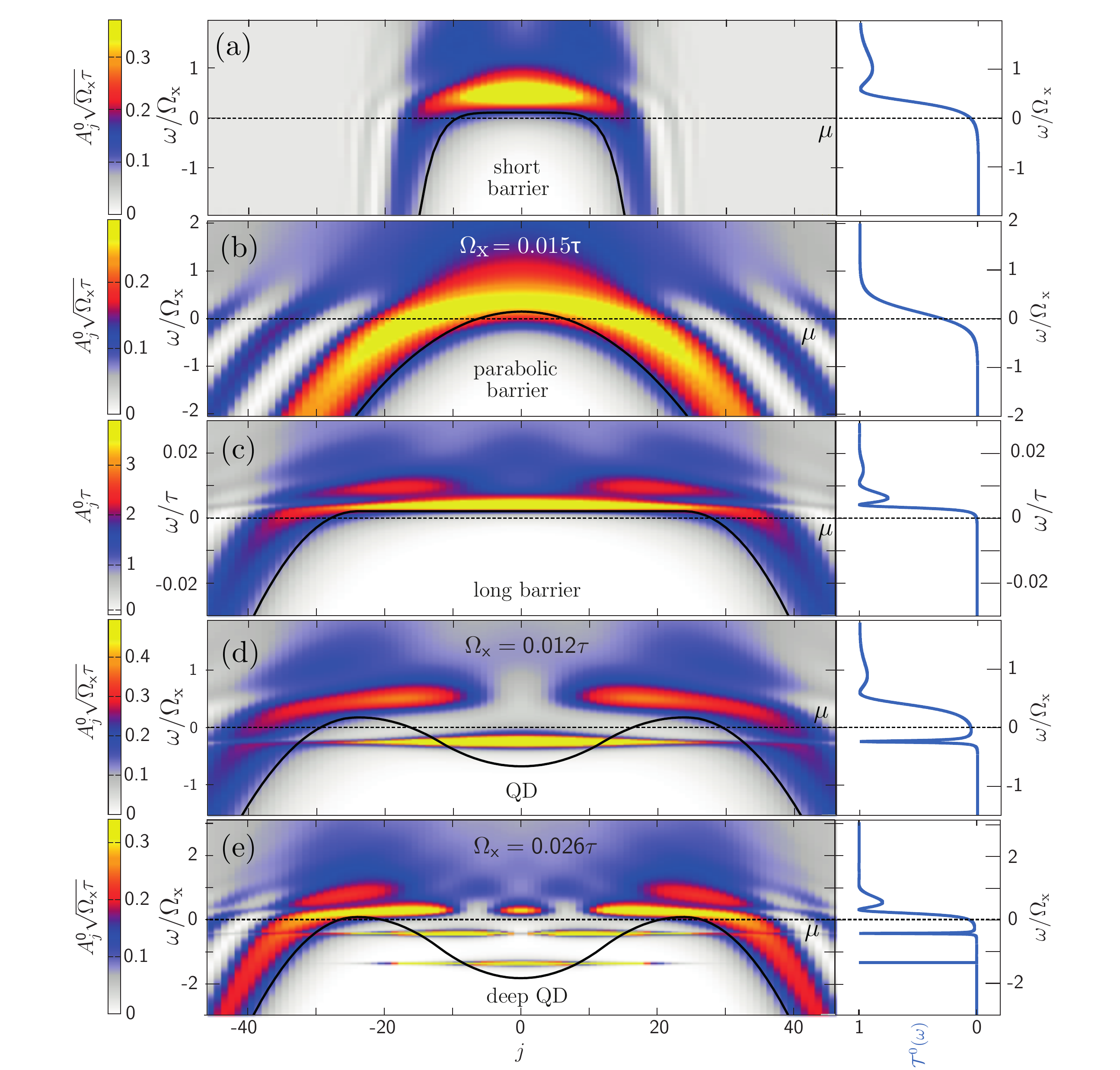}
  \caption{\label{fig:LDOS-QPC-QD} Noninteracting local density of
    states per spin species, $\dostateszero_j(\omega)$, and the
    noninteracting transmission per spin species, $\Tzomega$, for
    $\tB=0$. The energy $\omega$ is measured \wrt\ the chemical
    potential. The five panels show five potential barrier shapes
    occuring during the QPC-QD crossover, namely (a) a QPC with a
    short flat barrier, (b) a QPC with a parabolic barrier, (c) a QPC
    with a long flat barrier, (d) a shallow QD with just one discrete
    orbital state, and (e) a deeper QD with two discrete orbital
    states.}
\end{figure*}

To convey some intuition for the geometrical properties of this model
in the absence of interactions, Fig.~\ref{fig:LDOS-QPC-QD} shows the
noninteracting LDOS, $\Ajzo$, and the noninteracting transmission
probability, $\Tzomega$ (with $\omega$ measured \wrt\ the chemical
potential), for five different choices of the barrier shape, chosen to
represent various states of the geometric crossover between a QPC and
a QD.  For a parabolic QPC [\Subfig{fig:LDOS-QPC-QD}{b}], the LDOS
exhibits a broad ridge (yellow-red) just above the band bottom,
$\omegajmin = \Ej - \eFermi$ (solid black line), 
which follows the shape of the barrier. This is the van Hove
ridge mentioned in the introduction; it originates from the 1D van
Hove singularity at the band bottom, which in the CCR is smeared out
on a scale set by the barrier curvature $\Omegax$.  When $\omega$ is
increased from below to above the barrier top, set by $\tVc$, the
transmission $\Tzomega$ changes from 0 to 1 in the form of a smooth
monotonic step of width $\Omegax$, centered at $\omega=\tVc$.

Upon raising the side gate parameter $\tVs$ at fixed $\tVc$, the
effective barrier top eventually turns flat
[\Subfig{fig:LDOS-QPC-QD}{c}] and the ridge in the LDOS narrows (while
the maximal value of the LDOS above the barrier increases
accordingly).  This flatter-than-parabolic barrier shape causes the
noninteracting transmission $\Tzomega$ to show wiggles at the onset of
the $\Tzomega=1$ plateau, which we interpret as Fabry-Perot-like
resonances. They are discussed in more detail in the next
subsection. Experimentally, we also observe the case of a short flat
barrier for quite large $\Vs$, which we simulate in
\Subfig{fig:LDOS-QPC-QD}{a} by using a short barrier with a quartic
top (described by Eq.~(\ref{eq:non-parabolic-barriers}) below, with
$n=4$). In agreement with the measured $g(\Vc)$ in \Fig{fig:rawdata}
we observe Fabry-Perot-like resonances in $\Tzomega$ for both cases,
short versus long flat barriers in panels (a) and (c), and the period
of the wiggles is longer for the shorter barrier, as expected.

When the central gate parameter $\tVc$ is lowered below $\tVs$,
we enter the QD regime [\Subfigsand{fig:LDOS-QPC-QD}{d}{e}]. The LDOS
now develops bound states, very narrow in energy, that are spatially
localized inside the QD and define its single-particle spectrum. They
are accompanied by resonances in the noninteracting
transmission. Note, though, that the energy beyond which the
$\Tzomega=1$ plateau associated with full transmission sets in, is
still determined by the broader LDOS ridges above the tops of the left
and right barriers, which are remnants of the van Hove ridge found for
the parabolic and flat barrier shapes in (b) and (a,c),
respectively. This is clearly seen in the transmission curves in
panels (c) and (e), which exhibit very similar Fabry-Perot-like
resonances near $\Tzomega=1$.  In addition, $\Tzomega$ in panels (d,e)
shows sharp resonances at $\omega<0$, reflecting the bound states in
the LDOS. The occurrence of a conductance step together with sharp
resonances is a clear signature of the coexistence of a QD and a QPC;
our experimental data show corresponding features in the QD regime of
\Fig{fig:rawdata}.

In the outer flanks of the potential barrier, the LDOS has
  interference fringes with a period that scales as $1/v$, and the LDOS value
  averaged over several such fringes likewise scales as $1/v$, where
  $v_j(\omega)$ is the semiclassical velocity of an electron with
  kinetic energy $\omega - \omegajmin$ at site $j$. This explains the
  strikingly different behavior of the LDOS at the flanks of the
  potential maxima in \Subfigsand{fig:LDOS-QPC-QD}{a}{b}:
For the short, flat barrier
  with steep flanks in \Subfig{fig:LDOS-QPC-QD}{a}, the velocity
  $v_j(\omega)$ of electrons with $\omega \simeq \mu$ increases
  rapidly with $|j|$. As a consequence the LDOS at $\mu$ decreases
  rapidly and forms interference fringes with an
  correspondingly rapidly decreasing
  period. For the parabolic barrier of
    \Subfig{fig:LDOS-QPC-QD}{b} the flanks
    decrease much more slowly with increasing $|j|$, thus the
    corresponding increase in Fermi velocity, the decrease in the
    average LDOS and the decrease in the interference period all occur
    more slowly, too.

\subsection{Fabry-Perot resonances}
\label{sec:fabry-perot}

In this section, we discuss the Fabry-Pero-like resonances (wiggles)
that are seen in both the measured conductance in \Fig{fig:rawdata}
(marked by dashed arrows) as well as in the calculated $\Tzomega$,
\eg\ in \Subfigskomma{fig:LDOS-QPC-QD}{a}{c}, in more detail.

\begin{figure}[ht]
\includegraphics[width=\linewidth]{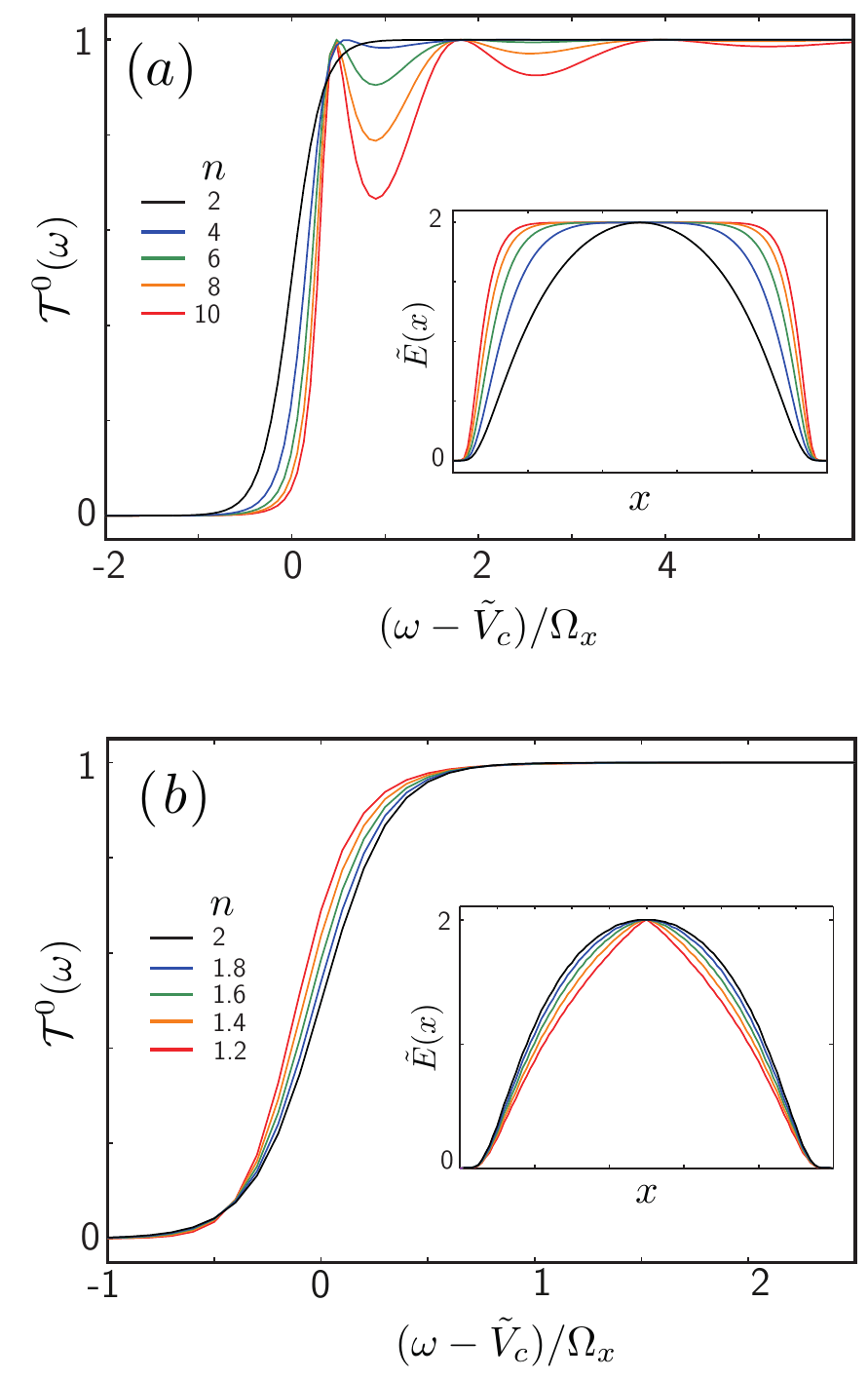}
\caption{Noninteracting transmission $T^0(\omega)$ as function of
  central gate voltage for several different barrier shapes,
  depicted in the insets, with barrier tops governed by $|x|^n$
    according to \Eq{eq:non-parabolic-barriers}. (a) Flatter-than
    parabolic barriers with $n\geq 2$, which arise during the QPC-KQD
    crossover; (b) sharper-than parabolic barriers with $n\leq 2$,
    shown for completeness.
    \label{fig:nonparabolic-barrier}}
\end{figure}

For our 1D model, studied in the absence of interactions, we find, in
particular, that $\Tzomega$ shows Fabry-Pero-like resonances whenever
the QPC barrier top is flatter than parabolic.  This is illustrated in
\Fig{fig:nonparabolic-barrier},
which shows the noninteracting
  conductance $g^0(\tVc)$ for a sequence of barrier shapes with
  barrier tops given by  
  \begin{eqnarray}
    \label{eq:non-parabolic-barriers}
  \tE (x) = \tVc + \eFermi - 
  \Omegax  \left(\frac{|x|}{l_x}\right)^n, \quad 
 l_x = \sqrt{\frac{2 \hbar^2}{m \Omegax}} \; , 
\end{eqnarray}
where $l_x$ is a characteristic length.  The noninteracting
transmission of a purely parabolic barrier top ($n$=2, black line) is
a smooth function of energy, given by \cite{Buettiker1990}
\begin{eqnarray}
\label{eq:baretransmission}
\Tzomega = [e^{2 \pi   (\tVc-\omega)/ \Omegax}+1]^{-1} \; . 
\end{eqnarray}
In contrast, making the barrier top flatter than parabolic by
  increasing $n$ introduces additional wiggles or resonances in
  $\Tzomega$, see \Fig{fig:nonparabolic-barrier}(a). Note that such
structures occur naturally in the conductance of longer QPCs and we
suspect that some previously published QPC measurements have likely
been performed in this regime of flatter-than-parabolic barriers
\cite{Iqbal2013,Brun2014}. Our own experimental results, displayed in
\Fig{fig:rawdata}, demonstrate that the transition from a gate defined
QD to a QPC likely covers the regime of a long QPC with a
flatter-than-parabolic barrier top and, moreover, a short QPC
with steep flanks can also result in a flatter-than-parabolic
barrier and Fabry-Perot-like resonances. 

For completeness, \Fig{fig:nonparabolic-barrier}(b) shows
  examples of $n\leq2$. Here, the transmission increases purely
  monotonically, without any Fabry-Perot-like resonances. 
  With decreasing $n$, the potential flanks tend to ``flatten'',
  causing the conductance step to develop an increasingly
  skewed shape: the step's onset becomes noticeably
  steeper, while the onset of the plateau is affected only weakly.

We note that it is not straightforward to distinguish Fabry-Perot-type
resonances, that occur even without interactions, from many-body
effects, that arise in the presence of interactions.  In the light of
recent experimental work on shape-dependent barriers, including Refs.\
\onlinecite{Iqbal2013,Brun2014} and this work, a systematic
theoretical study of how Fabry-Perot-type resonances are affected by
turning on interactions would be very interesting, but is beyond the
scope of this work.

\subsection{fRG approach}
\label{subsec:fRG}

To theoretically study the effect of interactions on the properties of
the CCR at zero temperature, we used
fRG\cite{Andergassen2006,Karrasch2006a,Metzner2012,Bauer2014}, a
renormalization-group-enhanced perturbative expansion in the
interaction. We used it to calculate the linear conductance $g$ of the
CCR, and three local quantities, the occupation $n_j$, magnetization
$m_j$ and spin susceptibility $\chi_j$ of site $j$, defined,
respectively, as
\begin{subequations}
  \label{eq:localquantities}
\begin{eqnarray}
  \label{eq:localdensity}
n_j & = & \langle \hat n_{j\uparrow} + \hat n_{j\downarrow}\rangle \; , 
\\
  \label{eq:localmagnetization}
m_j & = & \langle \hat n_{j\uparrow} - \hat n_{j\downarrow} \rangle/2 \; , 
\\
  \label{eq:localsusceptibility} 
\chi_j & = & \partial_\tB m_j|_{\tB=0} \; . 
\end{eqnarray}
\end{subequations}
The results are presented in Secs.~\ref{sec:magnetoconductance}
and \ref{sec:localproperties}, below.
%
%$n_j= \langle \hat n_{j\uparrow} + \hat n_{j\downarrow}\rangle$, the local
%magnetization $
%m_j= \langle \hat n_{j\uparrow} - \hat n_{j\downarrow} \rangle/2
%$, and the local spin susceptibility, 
%$\chi_j = \partial_\tB m_j|_{\tB=0}$ 

The details of our fRG approach are explained concisely in the
supplement of Ref.\ \onlinecite{Bauer2013}, and in more detail in
Ref.~\onlinecite{Bauer2014}. Here we just summarize some key
aspects. We restrict ourselves to zero-temperature calculations in the
Matsubara formalism.
% Calculations at finite temperature would require Keldish fRG and are
% beyond the scope of this article.  
Our fRG flow equations are based on two criteria. First, we assume
that $m_j = 0$ for $\tB = 0$, thus spontaneous symmetry breaking is
ruled out \emph{a priori}. This assumption is justified \emph{a
  posteriori} by the agreement of our fRG results with experiment,
both in Ref.\ \onlinecite{Bauer2013} and in the present paper.
Second, we neglect all contributions to the flow of the interaction
vertex that are not already generated to second order in the bare
(onsite) interaction, but feed back all other terms. This so-called
coupled latter approximation\cite{Bauer2014} amounts to including all
RPA-like channels on equal footing, while feeding back all
Hartree-like terms into the Fock-like equations and vice versa.  As a
computational simplification, we here use a ``static'' version of the
coupled-ladder approximation, which neglects all frequency
dependencies in self-energies and vertices. For the model of present
interest, the results for the zero-temperature conductance obtained
via this static simplification are qualitatively essentially the same
as those obtained by a ``dynamic'' calculation in which the frequency
dependence is retained, as shown explicitly in
Refs.~\onlinecite{Bauer2013,Bauer2014} for a parabolic QPC potential.

The effective expansion parameter for static fRG is $U_j
\dostateszero_j (0)$.  As a result, we find that the fRG equations
describing vertex flow do not converge for geometries that cause $
\dostateszero_j (\omega)$ to be sufficiently sharply peaked near the
chemical potential, i.\,e.\ near $\omega = 0$.  This problem occurs in
the QD regime, where the shallow few-electron QD has wide barriers
near $\mu$. We have therefore neglected vertex flow for plots that
involve this regime, i.\,e.\ in \Subfigsto{fig:QD-QPC}{a}{c}, and for
the movie discussed in App.~\ref{sec:movie}. Vertex flow was included,
however, for all other fRG results shown in this paper, and is
essential for obtaining the 0.7-shoulder in the QPC conductance even
at $\tB= \tilde T = 0$, as discussed in
\Sec{sec:magnetoconductance-QD-QPC} below.

In \bauer, we showed that our model and fRG
treatment of interactions are able to capture key elements of the
0.7-anomaly in a QPC in a qualitatively correct manner, including its
magnetoconductance. In the next section, we show that this is true
also for the Kondo effect in a KQD, and in fact for the entire QD-QPC
crossover.

\section{Magnetoconductance}
\label{sec:magnetoconductance}

In this section, we compare zero-temperature fRG results and
  low-temperature experimental data for the conductance during
  geometrical QD-QPC crossover, for three different magnetic fields
  [\Sec{sec:QD-QPC-crossover-finite-field}]. We also discuss the
  magnetoconductance in the KQD and QPC regimes in more detail,
  showing that fRG reproduces the characteristic magnetic-field
  dependence associated with the Kondo effect and the 0.7-anomaly,
  respectively [\Sec{sec:magnetoconductance-QD-QPC}]. Moreover, we
  discuss the $\tVc$-dependence of the characteristic low-energy scale,
  $\tBstar$, that characterizes the strength of the magnetoconductance
  at low fields, and its relation to the static spin susceptibility of
  the CCR [\Sec{sec:Bstar}].

\subsection{QPC-QD crossover at finite magnetic field}
\label{sec:QD-QPC-crossover-finite-field}

\begin{figure*}
\centering
  \includegraphics[width=\linewidth]{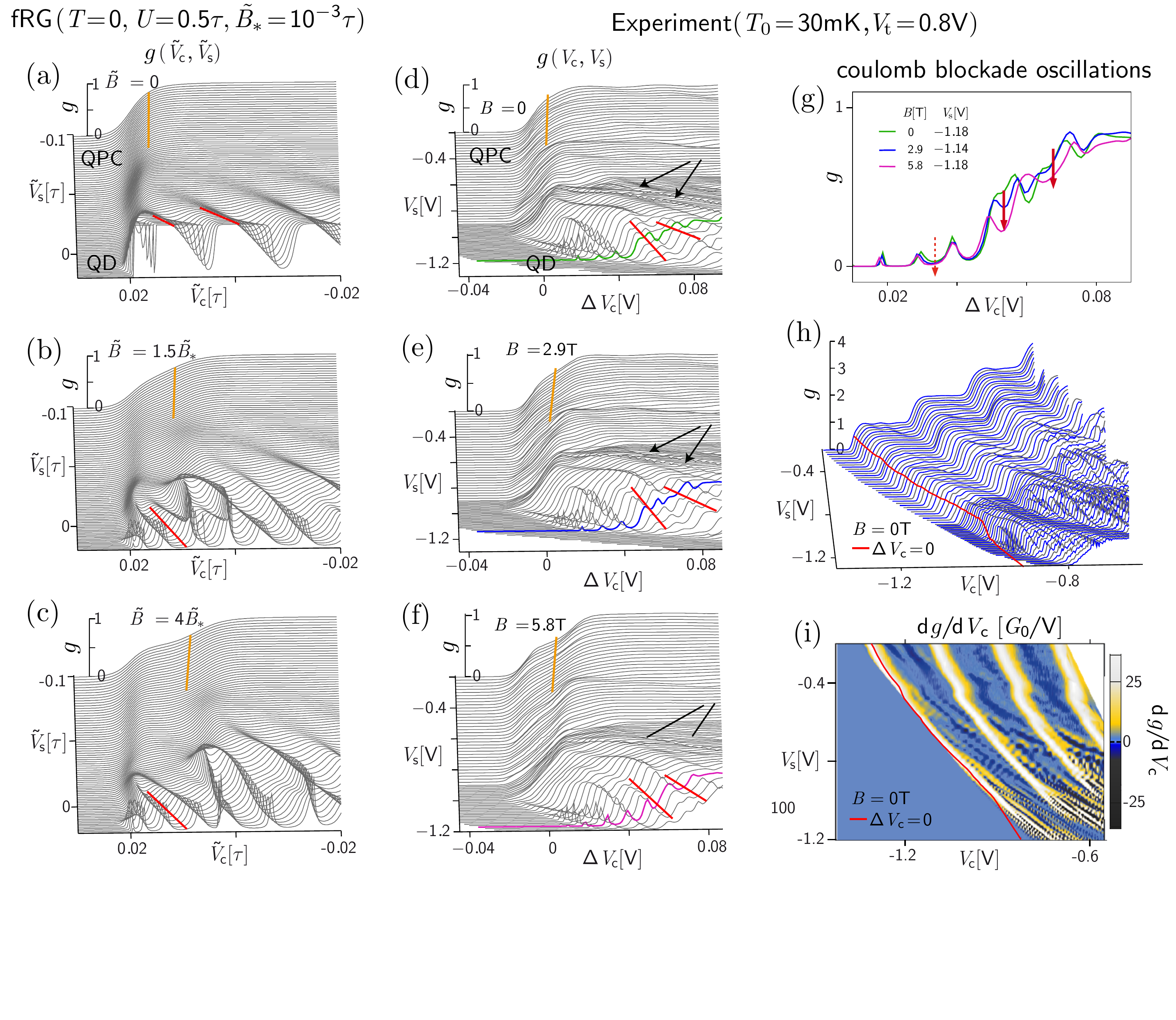}
  \caption{Transition from a QD to a QPC. (a-c) fRG1 results for the
    conductance $g(\tVc,\tVs)$, calculated at $T=0$ and three
    different fields, and plotted as function of the central gate
    voltage $\tVc$ for a large number of different side gate voltages
    $\tVs$.  (d-f) Analogous to (a-c), but showing experimental data
    for the conductance $g(\Vc,\Vs)$ in the range $0\le g\le 1$,
    measured at a fixed low temperature, $T_0 = 30$~mK. For each side
    gate voltage $\Vs$, the conductance trace is plotted as function
    of the shift $\Delta \Vc = \Vc - \Vcpinch$ in central gate voltage
    $\Vg$ relative to a $\Vs$-dependent reference value $\Vcpinch
    (\Vs)$, which is indicated by red lines in panels (h) and (i). [We
    chose $\Vcpinch$ to statisfy $g(\Vcpinch)=0.5$ in the QPC regime
    ($\Vs > -1$\~V), and to shift linearly with $\Vs$ in QD regime,
    with a slope chosen such that the red line does not cross any
    resonances.]  Orange lines in (a-f) mark the 0.7-anomaly, red
    lines mark Kondo valleys; black arrows in (d-f) mark Fano
    resonances.  (g) The three colored pinch-off curves from (d-f),
    all measured at $\Vs=-1.18\,$V, are plotted together to show how
    in Kondo valleys (marked by red arrows) the Kondo-enhanced
    conductance is suppressed by increasing field. (h) The raw
    experimental data for $g(\Vc,\Vs)$ at $B=0$ [corresponding to (d)]
    is plotted over a larger range of (unshifted) gate voltages to
    show several conductance steps. (i) 
    The derivative ${\rm d} g/{\rm d}\Vc$ of
    the data from panel (h). [(h) and (i) show identical data as Fig.\
    \ref{fig:rawdata} and its  inset.] The red lines in (h) and (i) show $\Vcpinch(\Vs)$, as used in (d-f).
    \label{fig:QD-QPC}}
\end{figure*}

\Fig{fig:QD-QPC} compares fRG results and experimental data for the
QD-QPC crossover at three magnetic fields, including the $B=0$ data
already shown in \Fig{fig:rawdata}.  \Subfigsto{fig:QD-QPC}{a}{c}
present model calculations of the zero-temperature linear conductance
$g(\tVc,\tVs)$ (using fRG without vertex flow) for three magnetic
field values, and \Subfigsto{fig:QD-QPC}{d}{f} corresponding
experimental data. The pinch-off value of $\Vc$ around which the
measured conductance drops to zero as $\Vc$ is decreased, say
$\Vcpinch$, is indicated using red lines in the raw data for $B=0$ in
\Subfigskomma{fig:QD-QPC}{h}{i}, which are miniature versions of
\Fig{fig:rawdata} and its inset. Note that $\Vcpinch$ shifts as a
function of $\Vs$, reflecting the capacitive influence of $\Vs$ on the
local potential between the center gates. This effect is absent in the
calculated data, since our model does not include such a
cross-coupling.  For better comparison between theory and experiment,
this cross-coupling is corrected for in the measured data in
\Subfigsto{fig:QD-QPC}{d}{f}, by plotting them as function of $\Delta
\Vc = \Vc - \Vcpinch$.

The measured transition from a QD to a QPC in
\Subfigsto{fig:QD-QPC}{d}{f} is smooth regardless of $B$. Our
calculations qualitatively reproduce the main features of the measured
QPC-QD crossover: Just as for the
  $B=0$ data in \Fig{fig:rawdata}, both the calculated and measured
  conductance traces in \Subfigsto{fig:QD-QPC}{a}{f} show the
  transition between a single QD with Coulomb
  blockade oscillations and a QPC with a smooth conductance step.
A movie showing how this crossover evolves
continuously with magnetic field is presented and discussed in
App.~\ref{sec:movie}.

Moreover, both the calculated (at $T=0$) and measured (at
$T_0\simeq30\,$mK) data exhibit the Kondo effect in the QD regime: it
manifests itself as an enhanced conductance in the Coulomb blockade
regime if an odd number of electrons charges the QD. In such Kondo
valleys, highlighted in \Subfigsto{fig:QD-QPC}{a}{f} by red lines, the
Kondo-enhanced conductance is strongly suppressed with increasing
field. \Subfig{fig:QD-QPC}{g} illustrates this for the measured data
by showing in a single panel the three colored pinch-off curves from
\Subfigsto{fig:QD-QPC}{d}{f}, taken for three comparable values of
side gate voltage $\Vs$.  (These three values, $\Vs = -1.18$, $-1.14$
and $-1.18$, are not all the same, because a random charge fluctuation
had occurred in the sample between the respective measurement runs,
shifting the potential landscape by a small but noticable amount.)
The solid red arrows in \Subfig{fig:QD-QPC}{g} mark the two Kondo
valleys corresponding to the red lines in
\Subfigsto{fig:QD-QPC}{d}{f}. The dashed red arrow in
\Subfig{fig:QD-QPC}{g} marks a third Kondo valley at a smaller
$\Vc$-value, where, however, the Kondo effect is already very weak,
since the coupling to the leads is so small that $\Tk < T$.

In the regime of a QPC defined by a parabolic barrier (small $\tVs$,
large $\Vs$), both measurements and calculations display the typical
magnetic field dependence of the 0.7-anomaly [marked by orange lines
in \Subfigsto{fig:QD-QPC}{a}{f}], namely the development from a weak
shoulder at $g\simeq0.7$ for $B=0$ to a pronounced plateau at a
reduced conductance for finite magnetic fields.

\subsection{Magnetoconductance of QPC and KQD}
\label{sec:magnetoconductance-QD-QPC}

In this subsection we compare theory and experiment in more detail,
for the magnetoconductance at two fixed values of side gate voltage,
for which the system forms a KQD or a QPC, respectively. For the QPC,
we have tuned the experimental system to have a smooth plateau at
$g=1$ without any Fabry-Perot resonances on the first conductance
plateau ($\Vs= -0.4$\,V, compare \Fig{fig:rawdata}), while we use a
parabolic barrier top for the theoretical calculations.

\begin{figure}[t]
\includegraphics[width=\linewidth]{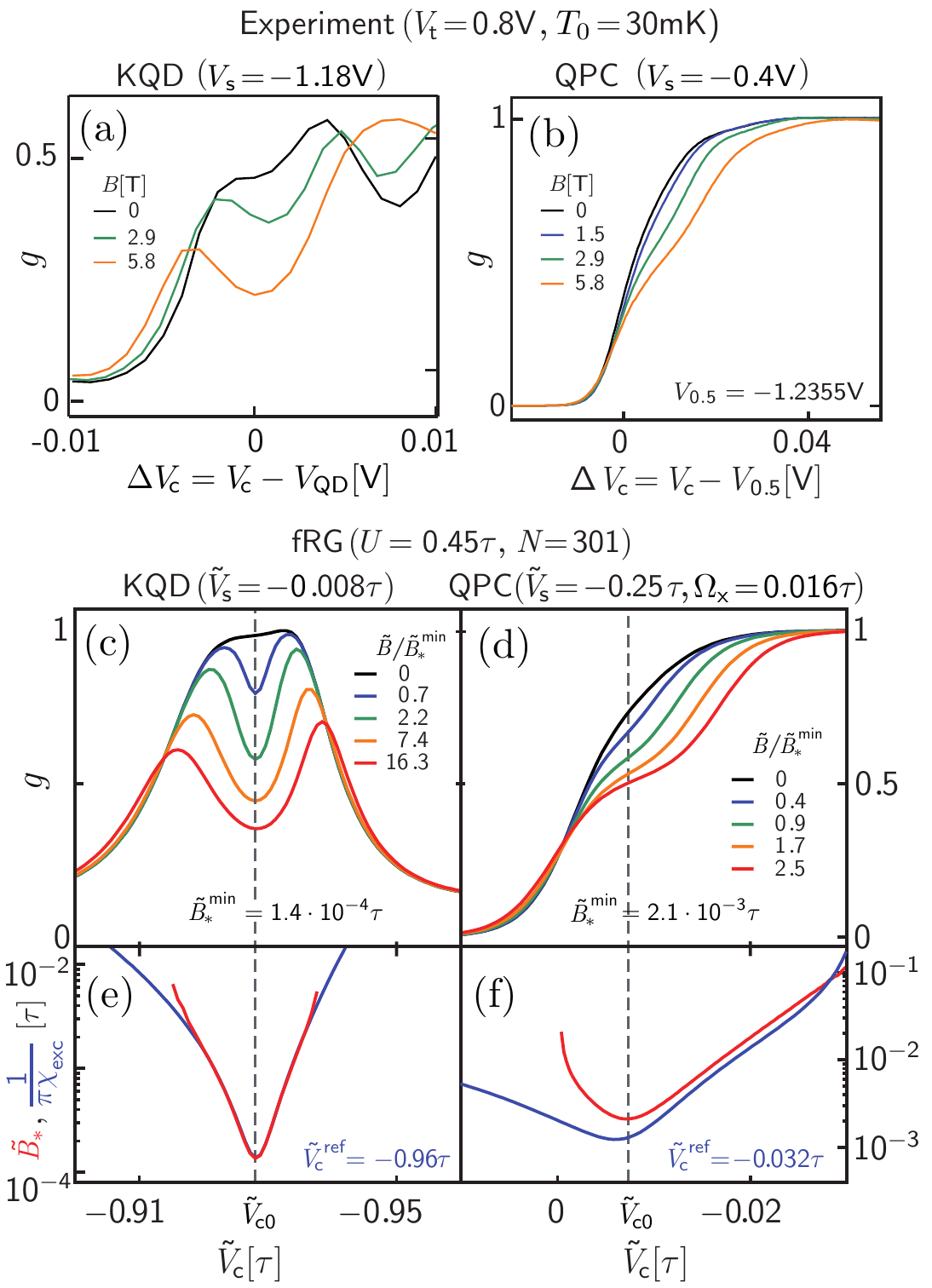}
\caption{Magnetoconductance: experiment vs.\ theory. (a,b)
  Experimental conductance curves for a KQD and a QPC, at two
  different, fixed $\Vs$-values and various magnetic fields measured
  at a low temperature $T_0=30$\,mK. Here, $\Delta \Vc$ is the offset
  of the central gate voltage $\Vc$ relative to $\VcQD$ or $\Vcpfive$,
  denoting the middle of the Kondo valley or the middle of the first
  conductance step, respectively. The data in (a) are a
  subsection of those shown in \Subfig{fig:QD-QPC}{g};
  the data in (b) correspond to those shown in \bauer,
  Fig.~2\textbf{e}.  (c,d) fRG results, plotted in a way analogous to
  (a,b), for the conductance $g(\tVc,\tB)$ at fixed $\tVs$ of a (deep)
  KQD containing $n^\KQD=49$ electrons, or for the lowest subband of a
  QPC, respectively.  (e,f) The corresponding KQD and QPC low-energy
  scales $\tBstar(\tVc)$ [red lines, from \Eq{eq:leading_h}] and
  inverse excess spin susceptibilities $1/[\pi \chiexc(\tVc)]$ [blue
  lines, from \Eq{eq:chibarrier}], plotted on a log-linear scale.
  Note that near the gate voltage $\tVczero$ [dashed line] where
  $\tBstar$ reaches its minimum, $\tBstarmin$, the small-field
  magnetoresponse in (c,d) is strongest. }
\label{fig:magnetocond}
\end{figure}
\Subfigskomma{fig:magnetocond}{a}{b} show measured conductance of a
KQD and a QPC, respectively, at several magnetic fields, $0\le
  B\le5.8\,$T, and \Subfigskomma{fig:magnetocond}{c}{d} show
corresponding fRG results (calculated with flowing vertex).  The fRG
calculations qualitatively reproduce the gate voltage and field
dependencies observed by us and numerous other experimental groups:
The conductance of the KQD [\Subfig{fig:magnetocond}{c}] shows a Kondo
plateau for $\tB = 0$, which is suppressed into a dip with increasing
field, as expected theoretically\cite{Costi2001,Karrasch2006a} and
observed experimentally in Ref.~\onlinecite{Nygard2000} and for our
own data [\Subfig{fig:QD-QPC}{g}, \Subfig{fig:magnetocond}{a}]. The
conductance step of the QPC [\Subfig{fig:magnetocond}{d}] exhibits a
0.7-shoulder at $\tB = 0$, which, as $\tB$ is increased, is suppressed
into a double step whose width is
  proportional to the magnetic field, as also seen in numerous
experiments\cite{Thomas1996,Cronenwett2002,Koop2007}, including our
own (\Subfig{fig:magnetocond}{b}, see also Ref.\
\onlinecite{Bauer2013}).

Note that the shoulder at $g\simeq0.7$ is visible in
\Subfig{fig:magnetocond}{d} even for $B=0$, much more so than in
\Subfig{fig:QD-QPC}{a} above; the reason is that the fRG scheme
without vertex flow used for \Subfigsto{fig:QD-QPC}{a}{c}
underestimates the effects of interactions compared to the fRG scheme
that includes vertex flow, used for
\Subfigskomma{fig:magnetocond}{c}{d}.  For a detailed discussion of
this point, see \bauer.

%   For both a KQD and a QPC the zero-temperature conductance depends
%   quadratically on magnetic field for small fields\cite{Bauer2013}:
% \begin{eqnarray}
%   \label{eq:Bstar}
%   g(\tVc,\tB) = g(\tVc,0) - \frac{B^2}{\tBstar(\tVc)} \; .
% \end{eqnarray}
% Here $\tBstar$ is a low-energy scale that quantifies strength of the
% magnetoconductance: the smaller $\tBstar$, the larger the
% magnetoconductance.  The dashed lines in
% \Subfigskomma{fig:magnetocond}{a}{b} show its $\tVc$-dependence on a
% log-linear scale. For a KQD, $\tBstar$ corresponds to the Kondo
% temperature, $\Tk$. It reaches its minimum near the center of the odd
% Kondo valley, in a way qualitatively consistent with the behavior
% known from the single-impurity Anderson model. For a QPC, $\tBstar$
% increases exponentially with $\tVc$ [straight dashed line in
% \Subfig{fig:magnetocond}{b}] in the gate voltage regime where
% $0.7 \lesssim g \lesssim 1$. As pointed out in
% \bauer, this dependence is an elementary
% consequence of the form of the bare transmission amplitude through a
% quadratic barrier.}

\subsection{Low-energy scale and excess spin susceptibility}
\label{sec:Bstar}

For both KQD and QPC, the low-field expansion of $g$,
\begin{equation}
\label{eq:leading_h}
g(\tB) \; \simeq \;
g(0) \bigl[1- (\tB/\tBstar)^2\bigr]
%_{\tB \rightarrow 0}
\quad (\tB \ll \tBstar)  \; ,
\end{equation}
can be used to characterize the strength of the $\tB$-dependence in
terms of a $\tVc$-dependent energy scale, $\tBstar$: the smaller
$\tBstar$, the larger the magnetoconductance.  For KQDs, the scale
$\tBstar$ in \Eq{eq:leading_h} corresponds to the Kondo temperature,
$\tBstar^\KQD = \kB \Tk$, according to Nozi\`{e}res' Fermi-liquid
\cite{Nozieres1974,Glazman2005} description of the low-energy limit of
the Kondo model.  The $\tB^2$-dependence (\ref{eq:leading_h}) has
recently been observed experimentally for a KQD\cite{Kretinin2011} and
previously for a few electron double quantum dot\cite{Schroeer2006};
for a QPC, it has been confirmed in \bauer\ Fig.~2\textbf{g} there).
Extracting $\tBstar (\Vc)$ from our fRG results
(\Subfigskomma{fig:magnetocond}{e}{f}, red lines) we find that for
both KQD and QPC it exhibits a distinct minimum, $\tBstarmin$, at
(say) $\tVczero$, near which it behaves as
\begin{subequations}
\begin{eqnarray}
\label{eq:T*-Vg-QD}
\widetilde B_\ast^\KQD & \propto & 
\exp \mbox{[}{c_1 (\tVczero - \tVc)^2 \mbox{]}} \; , 
% \ln \widetilde B_\ast^\KQD \jvd{-{\rm const.}}
%& \propto & (\tVczero - \tVc)^2
%  \; ,
\\
\label{eq:T*-Vg-QPC}
\widetilde B_\ast^\QPC & \propto & 
\exp \mbox{[}c_2 (\tVczero - \tVc)/  \Omega_x\mbox{]}
\; ,
% \ln \widetilde B_\ast^\QPC \jvd{-{\rm const.}} &  \propto &
% (\tVczero - \tVc)/  \Omega_x
\quad  (\tVc < \tVczero) \; , \qqph
\end{eqnarray}
\end{subequations}
for the KQD and QPC geometries, respectively.  ($c_1$ and $c_2$ are
$\tVc$-independent
constants.) %(with geometry-dependent proportionality constants,
\Eq{eq:T*-Vg-QD} reproduces for $\tBstar$ the behavior theoretically
predicted\cite{Haldane1978} and experimentally
observed\cite{Wiel2000,Kretinin2011} for the Kondo temperature of a
KQD.  The linear exponential behavior described by \Eq{eq:T*-Vg-QPC}
for a QPC is valid even in the absence of interactions. There it
follows directly from the non-interacting transmission formula for
parabolic barriers, \Eq{eq:baretransmission} (see Sec.~S-5
  of \bauer). Experimentally, \Eq{eq:T*-Vg-QPC} has been confirmed
in \bauer\ (Fig.~2\textbf{e} there).  Thus, our 1D model and fRG
treatement of interactions correctly capture the full $B$- and $\Vc$-
dependence of the conductance of both KQD and QPC, including the
exponential dependence of $\Bstar$ on $\tVc^2$ or $\tVc$,
respectively.

For the Kondo effect, the scale
$\tBstar$ defined by \Eq{eq:leading_h} is inversely proportional to
the excess contribution of the KQD to the static
spin susceptibility at zero temperature, 
\begin{eqnarray}
\label{eq:Bstarchi-universal}
1/\tBstar^\KQD = \pi  \chiexc. 
\end{eqnarray}
This relation, which links the strength
  of the magnetoconductance to that of local spin
  fluctuations, is a hallmark of Nozi\`eres's Fermi-liquid
  theory.\cite{Nozieres1974}  For our model, we define the excess spin
  susceptibility of the CCR by
\begin{eqnarray}
\chiexc (\tVc)
= \sum_{j \in \CCR} \left[\chisj (\tVc) -
\chisj (\tVcref)\right]\; ,
\label{eq:chibarrier}
\end{eqnarray}
where $\chisj$ [\Eq{eq:localsusceptibility}] is the local zero-field
spin susceptibility of site $j$, and $\tVcref$ a reference potential
at which the magnetoconductance is very small.  As reference for a
KQD, we take $\tVcref$ to define an even QD (EQD) charged by an even
number of electrons in an adjacent Coulomb-blockade valley; for a QPC,
we take $\tVcref$ small enough to define a truly open 1D channel ($g >
0.999$).  We find that the characteristic Fermi-liquid relation
\begin{eqnarray}
\label{eq:Bstar-propto-chiinverse}
1/\tBstar \propto \chiexc
\end{eqnarray}
is satisfied very well for the KQD for $\tVc$ near $\tVczero$
[\Subfig{fig:magnetocond}{e}], as expected. Remarkably, we find
  that for a QPC, too, a small $\tBstar$ goes hand in hand with a
  large $\chiexc$. In fact, by using $\tVcref$ as fit parameter, the
  inverse proportionality \Eq{eq:Bstar-propto-chiinverse} can be
  achieved for a QPC over a rather large range of gate voltages $\tVc
  \lesssim \tVczero$, as shown in \Subfig{fig:magnetocond}{f}.
{That the inverse relation between $\tBstar$ and $\chiexc$ also
  holds roughly for a QPC (though not as well as for a KQD, and
  requiring a fit parameter in the definition of $\chiexc$) is truly
  remarkable and constitutes one of the main theoretical results of
  this paper: the link between the \textit{magnetoconductance} and
  \textit{local spin fluctuations} that characterizes the Fermi-liquid
  regime of the Kondo effect, namely \Eq{eq:Bstar-propto-chiinverse},
  applies for the 0.7-anomaly as well.  This substantiates the
  argumentation, presented by us in \bauer, that the \sevenKS\ is a
  manifestation of the fact that a KQD and a parabolic QPC show
  similar Fermi-liquid behavior at low energies.}

Below, we further explore this link by theoretically studying the
relation between the magnetoconductance and local properties in more
detail, and for several different QD-QPC crossover trajectories.

\section{Local properties} 
\label{sec:localproperties}

In this section we study the local properties of the CCR at zero
  temperature, calculated by fRG.  We here focus on the local density
  $n_j$, magnetization $m_j$ and spin susceptibility $\chi_j$
  [defined in \Eq{eq:localquantities} above].
 We  find that a strong negative magnetoconductance goes 
hand in hand with an enhanced local spin susceptibility in the CCR, 
and argue that this connection is the microscopic origin
  of the \sevenKS.

% $n_j= \langle \hat n_{j\uparrow} + \hat
%  n_{j\downarrow}\rangle$, the local magnetization $m_j= \langle \hat
%  n_{j\uparrow} - \hat n_{j\downarrow} \rangle/2$, and the local spin
%  susceptibility, $\chi_j = \partial_\tB m_j|_{\tB=0}$.}
% The main conclusion  will be
%  that the \sevenKS\ is due  to an enhanced local spin susceptibility,
%  present for both a QPC and a KQD.

%
\begin{figure*}
\includegraphics[width=\linewidth]{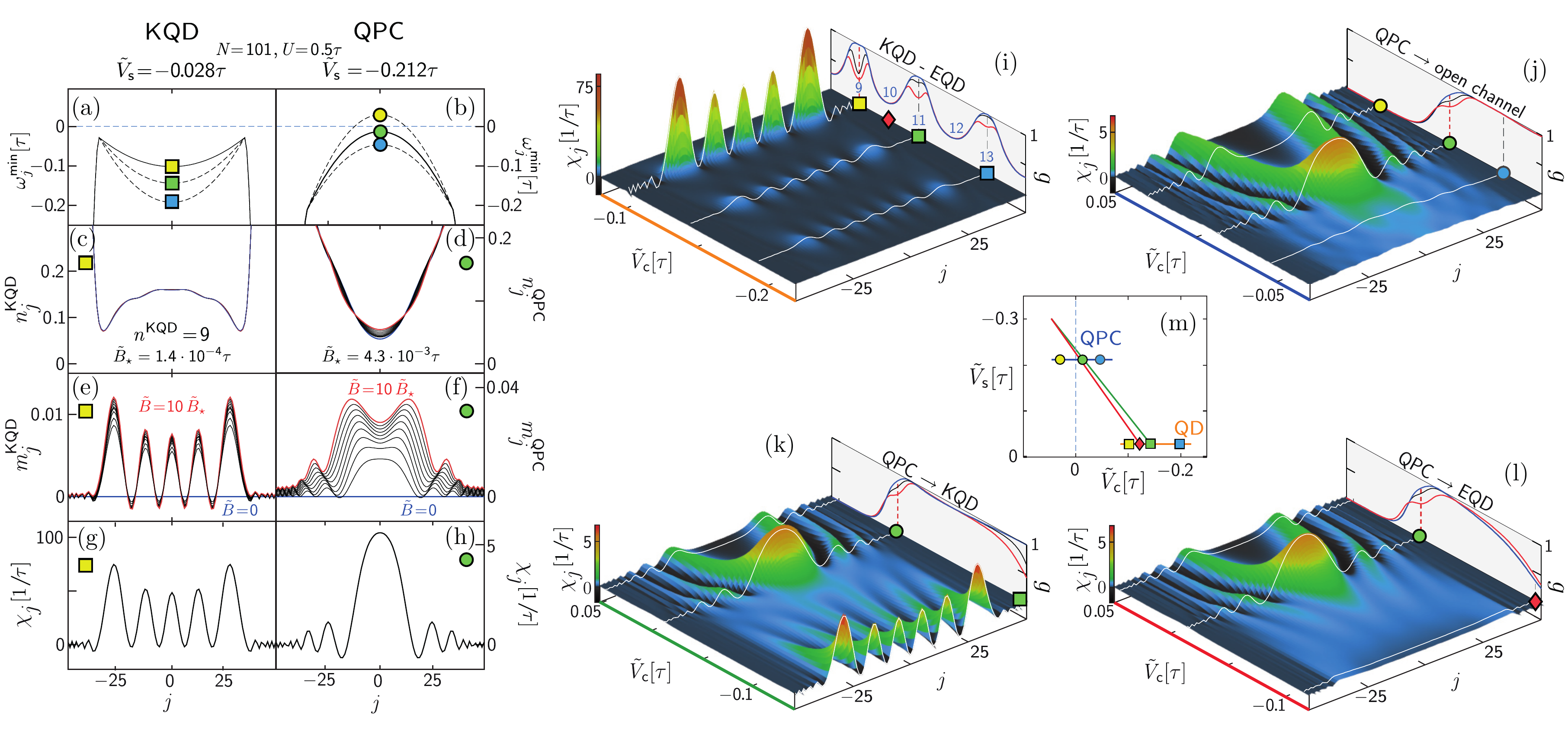}
\caption{\label{fig:local}\label{fig3} Local properties of KQD 
[charged with $n^\text{KQD}=9$ electrons for (c,e,g)]
  vs. QPC, calculated using fRG including vertex flow.  (a-f)
  \textit{Fixed geometries of KQD (left) and QPC (right)}: (a,b) The
  barrier tops for a (shallow) KQD and a QPC near pinchoff,
  respectively; for each, we show three barrier shapes $\omegajmin =
  \Ej - \eFermi$, used to calculate the curves in (c-l) marked by
  matching colored symbols.  (c,d) Local density $n_j$, and (e,f)
  local magnetization $m_j$, for 11 equidistant fields from $\tB = 0$
  (blue) to $\tB=10 \tBstar$ (red), for the KQD and QPC potentials
  shown by solid lines in (a,b), respectively.  (g,h) The local spin
  susceptibility $\chisj$ for the KQD and QPC potentials shown by
  solid lines in (a),(b), respectively; the spatial structure of
  $\chisj$ reflects that of $m_j$ for small fields, likewise showing
  strongly-enhanced standing-wave oscillations in the CCR.  (i-l)
  \textit{Geometric crossovers:} $\chisj$ is shown as a function of
  site $j$ and $\tVc=\widetilde V_{j=0}$ for four trajectories in the
  $(\tVc,\tVs)$ plane, drawn color-coded in panel (m) [where colored
  symbols mark $\tVc$- and $\tVs$-values used in (a-l)]. Each panel
  (i-l) also shows $g(\tVc)$ for three fields ($\tB/\tBstarmin = 0$,
  $\sim 1$, and $\gg 1$), to indicate the $\tVc$-dependence of the
  magnetoconductance; red dashed lines mark the gate voltage,
    $\tVczero$, where $\tBstar$ takes its minimal value,
    $\tBstarmin$.  (i) A QD being charged starting from 9 electrons (up left) to 13 electrons (down right), showing Coulomb blockade
  oscillations, (j) a QPC tuned from pinchoff to an open channel, (k)
  a crossover from a QPC to a KQD with 11 electrons, and (l) a crossover from a QPC to an even QD (EQD) charged by 10 electrons.
  \label{fig:chi}}
\end{figure*}
\Subfigsto{fig:chi}{c}{f} compare the $\tB$-dependence of $n_j$ and
$m_j$ of a KQD and a parabolic QPC near pinchoff, whose barrier shapes
are shown by solid lines in \Subfigskomma{fig:chi}{a}{b},
respectively.  Towards the edges of the CCR (large $|j|$, $\Ej \to
0$), the density $n_j$, plotted in \Subfigskomma{fig:chi}{c}{d}, rises
toward the filling of the non-interacting leads. For the KQD the
charge near the center of the CCR is well-localized and discrete
[$n_j^\KQD$ sums to $n^\KQD=9$ between the two distinct minima in
(c)].  For the parabolic QPC, in contrast, $n_j^\QPC$ is minimal at
the center, showing no signs of localized charge.  For $\tB \neq0$,
both $m_j^\KQD$ and $m_j^\QPC$, plotted in
\Subfigskomma{fig:chi}{e}{f}, show strongly-enhanced standing-wave
oscillations in the \CCR\ (with locally varying wavelength $\lambda
\sim 1/n_j$), but significant differences arise when $\tB$ increases
far beyond $\tBstar$: For a KQD, $m_j^\KQD$ saturates in magnitude,
its maxima stay fixed in position, and $n_j^\KQD$ remains
$\tB$-independent, all indicating that a discrete spin is being
polarized. In contrast, for a QPC, $m_j^\QPC$ \textit{does not
  saturate} for $\tB\gg\tBstar$, its maxima shift outward, and
$n_j^\QPC$ increases near the barrier center, all indicating that a
smooth redistribution of charge and spin occurs during the
polarization of the \CCR, which ultimately causes the spin-split
double conductance step at $\tB \gg \tBstar$.  We conclude that
whereas the KQD harbors a discrete, localized spin-$\frac{1}{2}$ local
moment, a parabolic QPC does not, since the spins in its \CCR\ are
neither discrete nor localized.  A detailed study of the behavior of
the magnetization in large fields $B>\Bstar$ follows in the next
section below.

Despite these differences, the KQD and QPC do show two striking
similarities in the regime of small fields, $\tB \ll \tBstar$,
relevant for the \sevenKS.  First, $m_j$ vanishes at $\tB=0$
(\Subfigskomma{fig:chi}{e}{f}, blue lines), reflecting our fRG
assumption that no spontaneous magnetization occurs, in contrast to
the spontaneous spin splitting scenario advocated in
Refs.~\onlinecite{Thomas1996,Thomas1998,Reilly2001,Smith2011} (see
\bauer, Supplementary Information, p.~5 and 6, for a detailed
discussion). Second, the local static spin susceptibility $\chisj$,
shown in \Subfigskomma{fig:chi}{g}{h}, exhibits a \textit{strong
  enhancement} (modulated by standing-wave oscillations) in the \CCR\
for \textit{both KQD and QPC}.  This enhancement arises through an
interplay of geometry and interactions. In the absence of
interactions, the bare local spin susceptibility in a QPC is
directly proportional to the LDOS at the chemical potential, $\chi_j^0
= \Ajzero/2$, and hence inherits the spatial dependence of the latter,
reflecting the geometry of the system. Interactions enhance the spin
susceptibility via a Stoner-type mechanism: upon turning on a small
Zeeman field that favors spin up over spin down, interactions enhance
the spin imbalance by further depleting the spin-down population.
The same line of arguments applies for a KQD in the low-energy
regime described by an effective Fermi-liquid Hamiltonian,
involving quasi-particles that experience a local interaction
whose strength is proportional to $1/\Tk$\cite{Nozieres1974}.

In contrast to a KQD, an EQD shows no $\chisj$-enhancement.  This is
illustrated by \Subfigsto{fig:chi}{i}{l}, which display $\chisj(\tVc)$
and $g(\tVc)$ for four trajectories in the $(\tVc,\tVs)$ plane,
corresponding to four types of geometric crossovers.
\Subfig{fig:chi}{i} shows a QD at fixed $\tVs$, whose electron number
(blue integers) is increased by lowering $\tVc$.  It exhibits odd-even
effects for both $\chisj(\tVc)$ and $g(\tVc)$: the Kondo-plateaus in
$g(\tVc)$ for odd electron numbers (KQDs) are accompanied by distinct
peaks in $\chisj(\tVc)$ (white lines) whereas the Coulomb valleys for
even electron numbers (EQDs) are not.  \Subfig{fig:chi}{j} shows a QPC
at fixed $\tVs$, which is tuned from pinchoff into an open channel
with $g=1$ by lowering $\tVc$.  The 0.7-anomaly in $g(\tVc)$ occurs
for $\tVc$ values near $\tVczero$ (red dashed line) where $\tBstar$ is
minimal. There the two maxima in $\chisj$ merge into a single one
(reminiscent of Figs.~2bA-C in Ref.~\cite{Rejec2006}), indicating that
the barrier top has dropped below 0 (compare panel
\Subfig{fig:chi}{b}, green circle), so that the chemical potential
cuts through the apex of the Hove ridge.  \Subfig{fig:chi}{k} shows a
QPC-KQD crossover ending in an 11-electron KQD: $\chisj$ exhibits
strong maxima in the QPC, which weaken in the open-channel regime
during the crossover to the KQD, where they become large
again. \Subfig{fig:chi}{l} shows a QPC-EQD crossover ending in a
10-electron EQD, where $\chisj$ remains very small, in contrast to the
case of the 11-electron KQD in \Subfig{fig:chi}{k}. Note that in the
QPC parts of \Subfigsto{fig:chi}{j}{l}, $\chi_j$ exhibits a
ridge-like, parabola-shaped main maximum as function of $\tVc$ and $j$
that mimicks (and indeed stems from) the Hove ridge in the LDOS as
function of $\omega$ and $j$ [\Subfig{fig:LDOS-QPC-QD}{a}].

The main message of \Subfigsto{fig:chi}{i}{l} is that the
\textit{negative magnetoconductance} seen for both KQDs and QPCs, but
not for EQDs, goes hand in hand with a \textit{strongly enhanced spin
  susceptibility}, whereas the latter vanishes or is weak for EQDs and
open 1D channels.  This is direct microsocopic evidence
  that the strong negative magnetoresistance observed
in both a KQD and a QPC as one of the key features of the \sevenKS,
originates from the fact that \textit{a QPC harbors
  strong local spin fluctuations} similar to those of a KQD. In this
regard, our scenario is fully consistent with the quasi-bound state
Kondo scenario proposed by Meir and collaborators
\cite{Meir2002,Golub2006,Rejec2006}. In fact, the spatial structure of
$\chisj^\QPC$ seen in \Subfigsto{fig:chi}{j}{l}, namely two peaks that
merge into one as $\tVc$ is lowered, is consistent with that of the
spin density of the ``quasi-bound states'' found for a QPC by SDFT
calculations \cite{Rejec2006,Ihnatsenka2007} (Figs.~2bA-C in
Ref.~\onlinecite{Rejec2006}). This is not surprising, since the SDFT
calculations were initialized using a small magnetic field to break
spin symmetry, which naturally gives rise to spin density maxima in
regions of large spin susceptibility.

We emphasize, though, that the \sevenKS\ applies only for low energy
scales, $\tB\ll \tBstar$, because while a KQD harbors a discrete,
localized spin-$\frac{1}{2}$ local moment, a QPC does not, as argued
above, and further elaborated in the next section. From the
perspective of the quasi-bound state scenario of Meir and
collaborators, this could be phrased by saying that the conditions for
the formation of a quasi-bound state cease to exist at large
fields. The differences between the 0.7-anomaly and the Kondo effect
are therefore evident in deviations of the QPC conductance from the
Kondo predictions as $T$ or $B$ approaches or exceeds $\Tstar$ or
$\Bstar$, as already detailed in \bauer.

\section{Magnetization}
\label{sec:magnetization}

In the previous section we have argued that the local magnetization
$m_j$ of a KQD and QPC evolve in strikingly different ways when $\tB$
increases far beyond $\tBstar$ [\SubFigsand{fig:local}{e}{f}]: For a
KQD, $m_j^\KQD$ saturates in magnitude, indicating that a discrete
spin is being polarized. In contrast, for a QPC, $m_j^\QPC$ shows no
signs of saturation, indicating that a smooth redistribution of charge
and spin occurs during the polarization of the \CCR. (Microscopically,
this originates from differences in the $\omega$-dependence of the
LDOS of a QD and QPC, illustrated in \Fig{fig:LDOS-QPC-QD} and
discussed in detail in Sec.~S-4.E of \bauer.) To substantiate our
  conclusion that a QPC does not harbor a discrete, localized
  spin-$\frac{1}{2}$ local moment, in contrast to a KQD, we present in
  this section additional fRG results on the evolution with $\tB$ of
  the magnetization, conductance and charge of a KQD and QPC. For
  comparison, we also include fRG results for the single-impurity
  Anderson model (SIAM), the paradigmatic model for local moment
  formation in metals.\cite{Anderson1961} It describes a local level
  with energy $\varepsilon_d = \tVc$ and Coulomb repulsion $U$ for
  double occupancy, that aquires a level width $\Gamma$ via
  hybridization with a conduction band of width $D$ (with $D\gg U \gg
  \Gamma$).

\begin{figure*}%[1]
  \includegraphics[width=\textwidth]{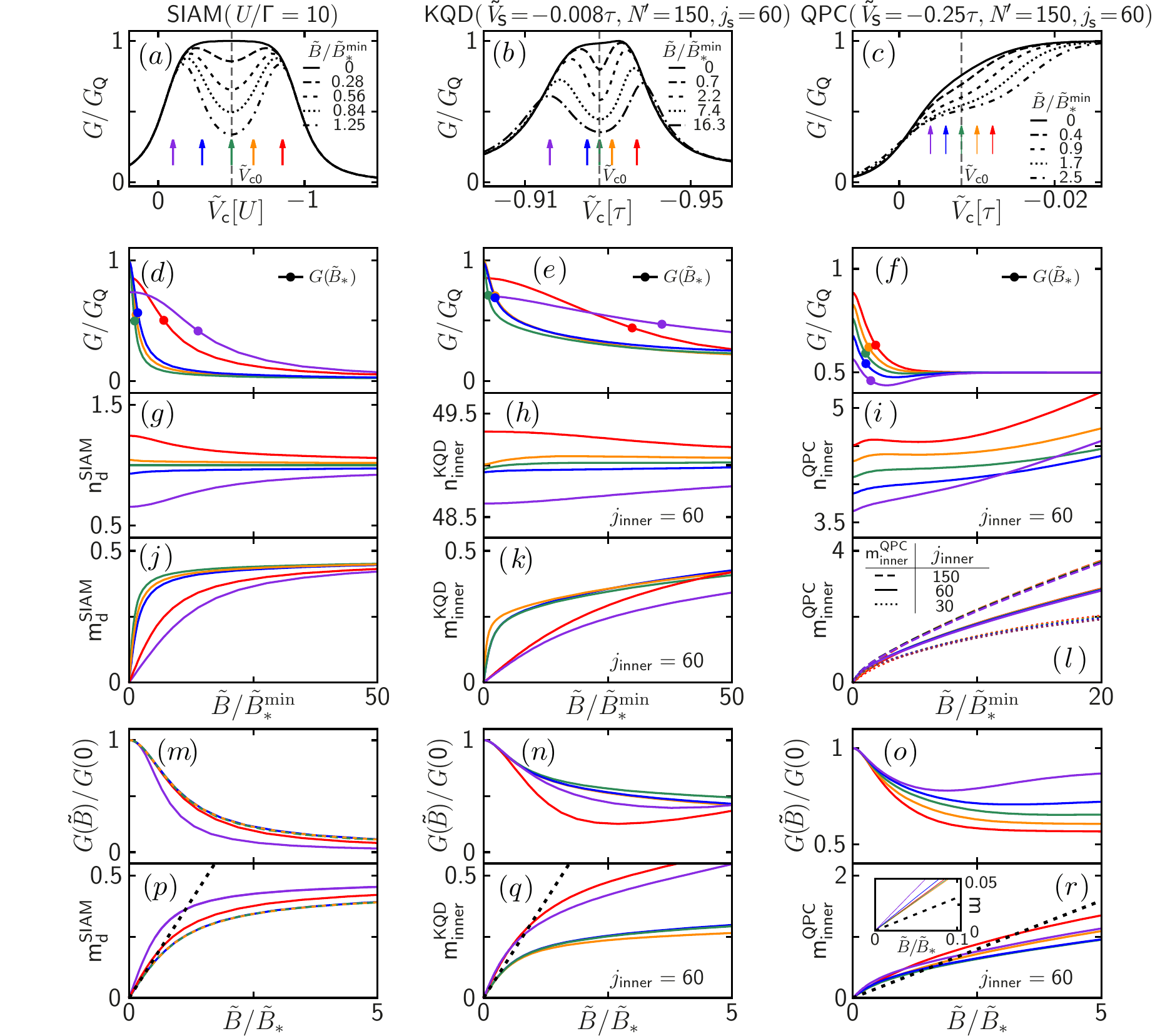}
  \caption{\label{fig:n_CCR-m_CCR-B-dependence} {\small fRG results
      (with vertex flow included) for the large-field behavior of
      the single-impurity Anderson model (SIAM, left column), a
      KQD (middle column, same parameters as
      \Subfig{fig:magnetocond}{c}, and a QPC (right column, same
      parameters as \Subfig{fig:magnetocond}{d}).  (a-c) The
      conductance $G$, plotted as function of $\tVc$, for five values
      of magnetic field; vertical dashed lines indicate the
      $\tVc$-value where $\tBstar$ is minimal, $\tBstar(\tVczero) =
      \tBstarmin$. (d-f) The conductance, (g-i) the total charge
      $\ninner$, and (j-k) the total magnetization $\minner$ in the
      CCR's inner region, comprising sites $|j| \le j_\inner$
      [\Eq{eq:define-spin-mCCR}].  These are plotted as functions of
      field $\tB/\tBstarmin$, for five different values of gate
      voltage $\tVc$, indicated by arrows of corresponding color in
      (a-c).  (For the SIAM, the CCR consists of just a single central
      site, which constitutes the local d-level of that model, thus
      $n_\inner^\SIAM= n_d^\SIAM$ and $m_\inner^\SIAM = m_d^\SIAM$.)
      (m-o) and (p-r) Same conductance and magnetization data as in
      (d-f) and (j-l), respectively, but plotted vs.\ $\tB/\tBstar$;
      black dotted lines in (p-r) have slope $1/\pi$, indicating the
      small-field limiting behavior $\minner = \tB / (\pi \tBstar)$
      expected in the Kondo limit [cf.\ \Eq{eq:Bstarchi-universal}].
      The inset of (r) shows a zoom of the limiting behavior for
      $\tB/\tBstar \to 0$.   }}
\end{figure*}
For this purpose, we define
the total charge and magnetization
 in the ``inner'' region of the \CCR\ by
%comprising the sights $|j| \le j_\inner$, by
\begin{eqnarray}
%\label{eq:define-charge-nCCR}
  \ninner  = \sum_{|j| \le j_\inner} n_j  \; , \quad
  \label{eq:define-spin-mCCR}
   \minner =  \sum_{|j|\le j_\inner} m_j  \; . \qqph
\end{eqnarray}
For the KQD geometry, we choose the inner region to lie between the
two maxima of the KQD potential, say at $\pm j^\KQD_\inner$.
% at $\pm j_s$ [cf.\ \Eq{eq:potential},
%\Subfig{fig:pot_expl}{c}], i.\,e.\ $j^\KQD_\inner = j_s$.  
The remaining CCR sites with $j^\KQD_\inner < |j| \le N'$ are
excluded, since they lie outside the dot, in the CCR barrier's outer
flanks.  Although the contribution of each such site to the CCR's
total charge or magnetization is small, their total contribution is
proportional to the length of the outer flanks, i.\,e.\ extensive, and
hence should be excluded when discussing intensive dot properties. For
the QPC geometry, in contrast, there is no natural separation between
an inner region and the barrier's outer flanks.  We will show results
for inner regions of three different sizes below: for fixed $N'=150$,
we choose $j^\QPC_\inner = 150$, 60 and 30 (they all yield
qualitatively similar results).

\Fig{fig:n_CCR-m_CCR-B-dependence} compares the large-field behavior
of the conductance, charge and magnetization of a KQD (middle column)
and a QPC (right column).  The left column shows corresponding
quantities for the SIAM, which corresponds to a CCR with just a single
site.  We denote its local charge and magnetization by $\ndSIAM$ and
$\mdSIAM$, respectively.

Panels (a-c) of \Fig{fig:n_CCR-m_CCR-B-dependence} show the
conductance as function of $\tVc$ for SIAM, KQD and QPC, respectively,
for five magnetic fields, specified in units of the $\tVc$-independent
reference field $\tBstarmin = \min[\tBstar(\tVc)]$ [cf.\
\Subfigskomma{fig:magnetocond}{c}{d}]. Colored arrows indicate five
fixed $\tVc$-values used to calculate the corresponding curves in all
other panels.  These show the conductance (panels
d-f), charge (panels g-i) and
magnetization (panels j-l) as functions of
$\tB/\tBstarmin$, as well as the scaled conductance $G(\tB)/G(0)$
(panels m-o) and magnetization (panels
p-r) as functions of $\tB/\tBstar$. For SIAM and
KQD, the blue, green and orange curves correspond to the local-moment
regime [$G/\GQ \simeq 1$ in (a,b), local charge close
to 1 in (g,h)], while the red and purple curves
correspond to the mixed-valence regime.

Upon comparing the three columns, we note the following salient
features: (i) For all five $\tVc$-values, the charges $\ndSIAM$,
$\ninnerKQD$ and $\ninnerQPC$ all depend only weakly on $\tB$
(g-i). In the small-field limit, the conductance and magnetization
shows Fermi-liquid behavior in all cases: (ii) the conductance
decreases quadratically with field (d-f), and (iii) the magnetization
increases linearly with field (g-i). (iv) At intermediate fields both
conductance (d-f) and magnetization (g-i) go through a crossover,
during which their slopes decrease markedly in magnitude.  (v) In the
large-field regime beyond this crossover, the behavior of the SIAM and
KQD differs strikingly from that of the QPC: SIAM and KQD exhibit
behavior characteristic of a spin-$\frac{1}{2}$ local moment, whereas
the QPC becomes spin-polarized with a magnetization much larger than
$\frac{1}{2}$.

To be specific, the evidence for this interpretation of the
large-field regime is as follows: (vi) For the SIAM, the plateau in
$\mdSIAM$ saturates towards $0.5$ (j); this saturation is the
hallmark of a polarized spin-$\frac{1}{2}$ local moment. (vii) At the
same time, the conductance $G$ continues to decrease with field,
albeit very slowly (d). (For the SIAM this decrease is known
to be logarithmic, $\sim 1/\ln(\tB/\tBstar)$, but fRG is not
sufficiently accurate to reproduce purely logarithmic behavior.)  The
KQD exhibits qualitatively similar features, though with some
quantitative differences: (viii) The plateau in $\minnerKQD$ is fairly
flat, too (k), although it does not truly saturate but
instead slowly increases past 0.5 for sufficiently large fields. This
reflects the fact that the KQD in \Fig{fig:n_CCR-m_CCR-B-dependence}
harbors not only one spin-$\frac{1}{2}$ local moment but many
additional occupied levels ($n_\mathrm{inner}\simeq50$); some of these
begin to contribute to the magnetization when $\tB$ becomes a
sufficiently large fraction of the dot level spacing.  (ix) The KQD
conductance continues to decrease with field (e), but less
slowly so than for the SIAM (d), due to contributions from
the additional levels.

The above large-field features of the SIAM and a KQD stand in stark constrast
to those of a QPC: (x) its magnetization continues
to increase with field without any saturation (l) [the slope
depends on the width of the inner region: the larger $j_\inner$, the
larger the slope]; and (xi) the conductance fully saturates at $G=
0.5\GQ$ (f), corresponding to a spin-split conductance
plateau. The absence of any saturation in the magnetization reflects
the fact that the QPC barrier lacks the isolated ``inner region'' of a
KQD. Instead, the CCR barrier is made up entirely of outer flanks,
along which electrons of both spin species can freely move. As the
magnetic field is increased, the magnetization of the QPC is thus free
to increase without any intrinsic limit (in contrast to the case of a
KQD). The spin-split conductance plateau sets in once the LDOS at the
chemical potential is fully spin-polarized.

These differences between SIAM and KQD on the one hand and QPC on the
other of course imply different behaviors when the conductance and
magnetization are plotted versus $\tB/\tBstar$: (xii) For the
local-moment curves (blue, green, orange) of the SIAM, the scaled
conductance $G(\tB)/G(0)$ (m) and the magnetization (p) both collapse
onto a single scaling curve when plotted versus $\tB/\tBstar$. (xiii)
The same is true approximately for the KQD's conductance (n) and
magnetization (q), though the collapse is not as perfect.  Thus, for
the SIAM and KQD, the $\tVc$-\textit{dependent} scale $\tBstar$
governs both the small- \textit{and} large-field behavior of the
magnetization and conductance. (xiv) This is not the case for the QPC,
whose conductance (o) and magnetization (r) do \textit{not} show a
collapse onto a single curve when plotted versus
$\tB/\tBstar$. (xv) Instead, the large-field behavior of the
magnetization is governed by a $\tVc$-\textit{in}dependent scale: when
the $\minnerQPC$-curves are plotted vs.\ $\tB/\tBstarmin$, they all
overlap (l), except in the limit $\tB \ll \tBstarmin$ [not resolved in
(l)]. Also, (xvi) the field scale at which the conductance saturates
at $G= 0.5 \GQ$ does not depend on $\tBstar$ at all, but instead grows
linearly with decreasing $\tVc$ (f).

To summarize: when the SIAM and the KQD are tuned into their local
moment regime, their conductance and magnetization exhibit
the expected crossover, governed only by a single energy scale
$\tBstar(\tVc)$, between a Fermi-liquid and a local-moment fixed point
that is characteristic of the Kondo effect, (xii,xiii). The QPC conductance
 and magnetization, however, do not, (xiv-xvi).  This is an example, therefore,
where the analogy between Kondo effect and 0.7-anomaly breaks down --
at large fields, they are distinct physical effects.

The lack of local-moment behavior for the QPC magnetization at large
fields is not surprising, given the open nature of the QPC geometry.
Nevertheless -- and this \textit{is} surprising and remarkable --
%the small-field limit ($\tB \ll \tBstar$) of $\minnerQPC$ \textit{is}
%governed by the $\tVc$-dependent scale $\tBstar$: in the limit
%$\tB/\tBstar \to 0$ (inset of panel \textbf{l}), those magnetization
%curves for which $\tVc < \tVczero$ have very similar slopes $\partial
%\minnerQPC/\partial \tB$, as is the case for $\partial \dSIAM/\partial
%\tB$ and $\partial \dSIAM/\partial \tB$, for which this slope is
%$1/\pi$ (dashed black lines in \textbf{j}, \textbf{k}).
(xvii) the small-field limit ($\tB \ll \tBstar$) of the magnetization
is governed by $\tBstar$ not only for the SIAM and KQD in the local
moment regime, but also for the QPC when $\tVc < \tVczero$: For all
these, the \textit{linear} response of the magnetization to field is
proportional to $1/\tBstar$, meaning that curves of $m_\inner$ vs.\
$\tB/\tBstar$ for different $\tVc$-values all have the same slope as
$\tB/\tBstar \to 0$.  [This is illustrated by the blue, green, orange
lines in panels (p) and (q), which all have slope $1/\pi$ (dashed
black line), in accord with \Eq{eq:Bstarchi-universal}; and by the
green, orange and red curves in the inset of panel (r), which have
mutually similar slopes, though these do not equal $1/\pi$ (dashed
black line).]  The fact that the small-field limit of the QPC
magnetization is governed by $\tBstar$, (xvii) has far-reaching
consequences, in that it underlies the low-energy Fermi-liquid
behavior of the QPC conductance mentioned in \Sec{sec:Bstar} above.

We end this subsection with a parenthetic remark: As an alternative to
\Eq{eq:define-spin-mCCR}, outer flank contributions to the
magnetization can also be eliminated by considering
\begin{eqnarray}
\label{eq:define-excess-m}
\mexc (\tVc) & = & \sum_{j \in \CCR} m_j (\tVc) - \sum_{j \in \CCR} m_j (\tVcref) \; , \qqph
\end{eqnarray}
the excess magnetization of the CCR at central gate voltage $\tVc$
relative to its magnetization at a suitably chosen reference voltage
$\tVcref$ [chosen to define an even QD (EQD) in an Coulomb blockade
value adjacent to the odd KQD, or an open QPC, as discussed in
conjunction with \Eq{eq:chibarrier}].  Indeed, for small fields
$(\tB/\tBstar \ll 1$) one finds $\mexc(\tB) \simeq \minner(\tB)$ when
choosing $j_\inner = j_s$, and the excess susceptibility defined in
\Eq{eq:chibarrier} corresponds to $\chiexc = (\partial \mexc
/ \partial \tB)_{\tB = 0}$.  However, for the large-field regime of
interest in the present subsection, the subtraction scheme of
\Eq{eq:define-excess-m} is not convenient,
% than for the small-field limit considered in the main article,
because at sufficiently large fields the second term becomes
comparable in size to the first, causing $\mexc$ to decrease.

\section{Summary and Outlook}
\label{sec:conclusions}

In \bauer, we have argued that the \sevenKS, \ie\ the observed
similarities in the low-energy behavior of the conductance for
0.7-anomaly and the Kondo effect, originate from geometry-induced,
interaction-enhanced local spin fluctuations, that are present both in
a QPC and a KQD. The goal of the present work has been to offer
additional evidence for this conclusion, by studying the geometric
crossover between a QD and a QPC, both experimentally and
theoretically, focusing on the magnetic field dependence at low
temperatures. Our experimental and numerical results were found to be
in good qualitative agreement. This shows that the 1D-model with
short-range interactions introduced here, together with the fRG
approach used to treat interactions, succeeds in capturing the
essential physics of the Kondo effect, the 0.7-anomaly and the
geometric crossover between them.

Our initial motivation for studying the geometric crossover was the
expectation that this would allow us to observe an adiabatic
transition from Kondo correlations present in a KQD to the
correlations present in a QPC showing the 0.7-anomaly.  Indeed, this
idea turned out to be fruitful: our fRG results show that an
anomalously strong negative magnetoconductance, one of the key
  features of the \sevenKS, always goes hand in hand with strongly
enhanced local spin fluctuations.   

 The spatial structure of the local spin fluctuations
is inherited from that of the non-interacting local density of states
at the chemical potential, and enhanced by interactions in Stoner-type
fashion. Roughly speaking, local spin fluctuations are strong in those
regions of space where the electrons near the chemical potential are
``slow''. For a parabolic QPC, slow electrons are found in the CCR
when the barrier top is just below the chemical potential, whereas for
a KQD the odd electron is slow simply because it is really trapped
inside the dot. The difference between these two situations does not
matter much for low energies ($\ll |\tBstar|$), thus both
show behavior characteristic of a Nozi\`eres-Fermi liquid
with local interactions. (These local interactions
are the reason why the \sevenKS\ also comprises similar
  temperature and source-drain voltage dependences 
for a KQD and a QPC, see \bauer\ for a more detailed
  discussion of this point.)
The difference does matter, though, for high energies, where we find
no indications that a parabolic QPC harbors a localized state, and
where indeed no \sevenKS\ is observed.

One of the lessons learnt from \Subfigsto{fig:local}{i}{l} is that
the presence or absence of the two crucial properties discussed
above, namely a strong negative magnetoconductance and strong local
spin fluctuations, depends very much on the trajectory followed in
the $(\tVc,\tVs)$ plane during the QD-QPC crossover. For example,
for the trajectory studied in \Subfig{fig:local}{k}, \textit{both}
these features disappear in the intermediate regime between the KQD
and the QPC, because there the barrier top is so far below the
chemical potential that the system is essentially an open channel,
with $g \simeq 1$.

It is, of course, possible to also implement QD-QPC crossover
trajectories during which the barrier top always remains close to the
chemical potential. Suppose that such a trajectory includes a wide,
flat barrier top, such as that shown in
\Subfig{fig:LDOS-QPC-QD}{b}. When this barrier top is just below the
chemical potential, the electron density will be low throughout the
wide barrier region, implying that interaction effects will become
very strong there. This regime is conducive to the formation of a
Wigner crystal, so that the conductance can be expected to show
behavior different from that of a ``standard'' 0.7-anomaly. Indeed two
recent experimental papers have studied this
regime\cite{Iqbal2013,Brun2014} and reported interesting differences
from standard 0.7-phenomenology (such as a zero-bias peak that splits
into two or even three subpeaks as the barrier width is varied).

In our own detailed studies of QPCs, both in \bauer\ and here, we
have so far purposefully chosen to avoid the regime of wide, flat
barrier tops. Instead, we have focussed on parabolic barrier tops
and demonstrated that these were sufficient to explain numerous
features of the standard 0.7-anomaly.  Nevertheless, it would be
very interesting to systematically study the crossover from
parabolic to wide, flat barrier tops.  The latter lead to
Fabry-Perot resonances even in the absence of interactions (as
argued in \Sec{sec:fabry-perot}), and the way in which Fabry-Perot
structures in the density of states are modified or enhanced by
interactions has not been explored systematically yet.

In the limit of a very wide and flat barrier, the CCR would represent
a long 1D wire of low density, behaving as a spin-incoherent Luttinger
liquid\cite{Matveev2004}. Since interaction effects become ever more
important as the density decreases, fRG will at some point become
unsuitable for a flat barrier top when either its width is made
sufficiently wide or its top approaches the chemical potential
sufficiently closely from below. However, more powerful
numerical methods, such as the density matrix renormalization group,
could be used to study such situations.  

%\papertwo{In a follow-up paper\cite{Delft2014}, we will explore
%the similarities and differences in the behavior of a KQD and a QPC
%in yet more detail. We there make a detailed comparative study of
%the transmission phase shifts through a KQD and a QPC, aiming to
%outline to what extent a future Nozi\`eres-Fermi-liquid theory for a
%QPC would differ from that of a KQD, and of the large-field
%magnetization of both.}

\section*{Acknowledgements} 

We thank B. Altshuler, P. Brouwer, R. Egger, J. Folk, L. Glazman,
V. Golovach, A. H\"ogele, Y. Imry, M. Kiselev, D. Logan, D. Loss,
C. Marcus, Y. Meir, M. Pustilnik, A. Rosch, H. Sellier,
K. Sch\"onhammer, B. Spivak and A. Yacoby for stimulating discussions,
and in particular S. Andergassen, C. Honerkamp, S. Jakobs,
C. Karrasch, V. Meden, M. Pletyukhov and H. Schoeller for fRG-related
help and advice.  We also thank Christoph Hohmann for producing
Figs.~1(b),(c).  We acknowledge support from the DFG via SFB-631,
SFB-TR12, De730/3-2, De730/4-1, De730/4-2, De730/4-3, HO 4687/1-3,
LU819/4-1, and the Cluster of Excellence \emph{Nanosystems Initiative
  Munich}, and the National Science Foundation under Grant No.  NSF
PHY05-51164. S.L. acknowledges support via a Heisenberg fellowship of
the DFG.

\appendix

\section{Evolution  of $g(\tVc,\tVs)$ with $\tB$ (movie)}
\label{sec:movie}

% \subsection*{Evolution  of $g(\tVc,\tVs)$ with $\tB$ (movie)}
In \Sec{sec:QD-QPC-crossover-finite-field}, we showed fRG results
(without vertex flow) for the QD-to-QPC crossover of the conductance
$g(\tVc,\tVs)$ as function of central and side gate voltage, for three
values of magnetic field [\Subfigsto{fig:QD-QPC}{a}{c}].  Its
continuous evolution with $\tB$, again calculated by fRG without
vertex flow, can be viewed as a QuickTimas a separate figuras a
separate figure Movie, see the file ``fRG.mov'' in the Supplementary
Material.  The movie shows simultaneously the evolution with $\tB$ of
three data sets: The central panel gives the conductance
$g(\tVc,\tVs)$ in a three-dimensional plot formatted in the same way
as \Subfigsto{fig:QD-QPC}{a}{c}. The top left panel gives the
frontmost curve of the central panel, $g(\tVc,\tVs = 0.018 \tau)$,
representing the pure QD regime; and the top right panel gives its
backmost curve $g(\tVc,\tVs = 1.9 \tau)$, representing the pure QPC
regime. A moving horizontal line in the scale bar on the right hand
side indicates the evolution of $\tB$, and whenever it passes one of a
selected set of $\tB$ values, that value is indicated by a frozen
horizontal line, while two curves of matching color freeze in the top
left and top right panels.

The initial evolution for small fields ($\tB \lesssim 2\times 10^{-4}
\tau$) shows how the Kondo plateaus of the first few Kondo valleys,
whose typical Kondo temperatures increase with dot occupancy $n$,
successively get suppressed as $\tB$ increases [see top left panel].
For larger fields ($\tB \gtrsim 2\times 10^{-4}$), the conductance in
the QPC regime also begins to develop a shoulder [see top right
panel], which evolves (beyond $\tB \gtrsim 3\times 10^{-3}$) into an
ever more pronounced double step.  Note that the scale bar changes
from logarithmic to linear at $\tB \simeq \times 10^{-4}$, since the
$\tB$-dependence of the conductance at large fields is logarithmic for
the Kondo effect, but linear for the 0.7-anomaly. (This is another
indication that the latter does not involve local-moment physics at
large fields.) 

\begin{figure}[t]
\centering
  \includegraphics[width=\linewidth]{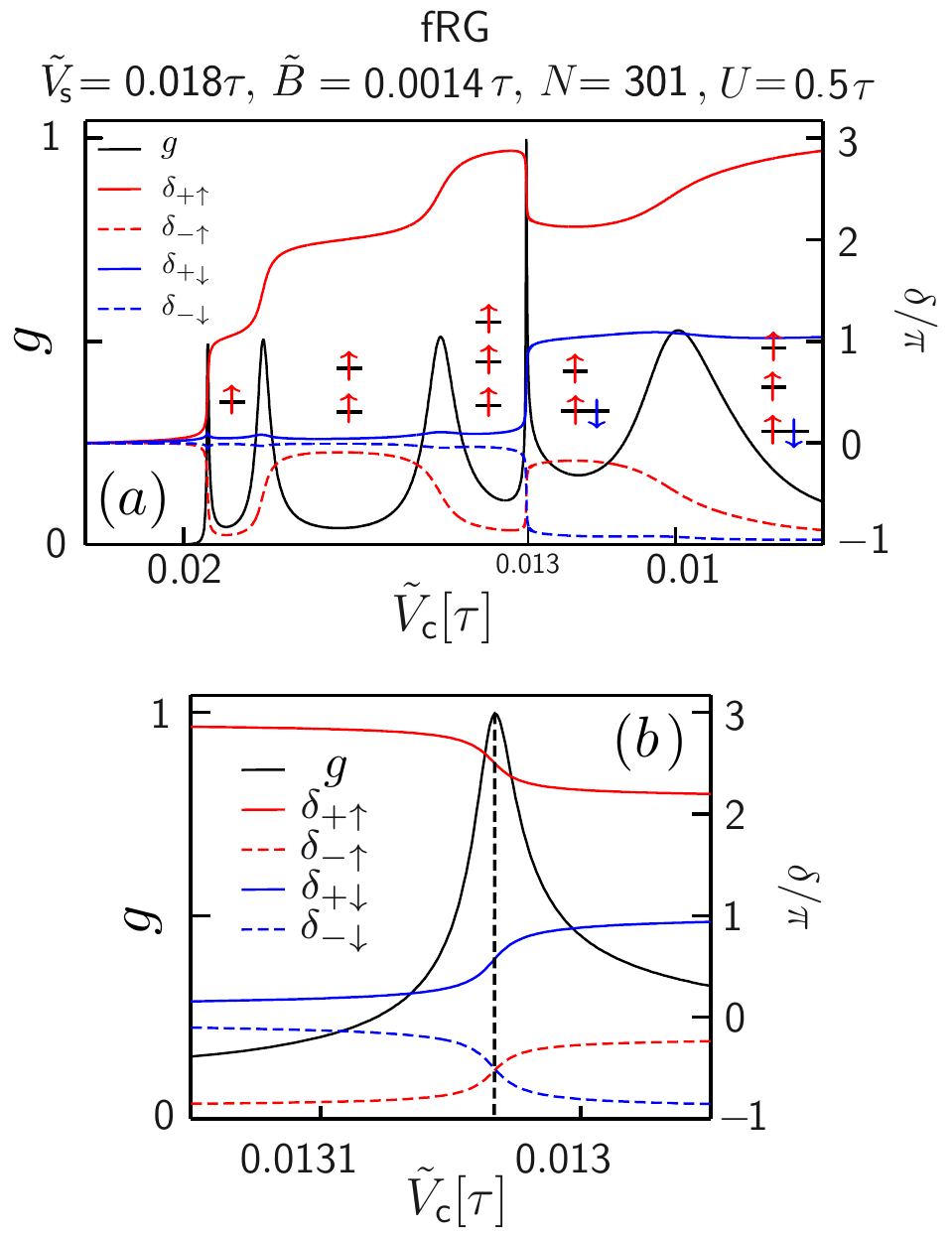}
  \caption{\label{fig:s1s3} Example of a spin-flip resonance in the
    conductance of a QD. (a) Conductance (solid black line) and phases
    $\delta_{\sigma,+}/\pi = n^\QD_\sigma$ (solid lines) and
    $\delta_{\sigma,-}/\pi$ (dashed lines), for $\sigma \! =\uparrow$
    (red lines) and $\sigma \! =\downarrow$ (blue lines), all
    calculated using fRG. The phases are all set to 0 at a reference
    voltage $\tVcr= 0.022$ so large that the \CCR\ no longer
    represents a QD, but a pinched-off QPC, with $g=0$ and
    $n_\sigma^\QD=0$.  As $\tVc$ decreases below $0.02\tau$ where the
    dot is still empty, $n_\uparrow^\QD$ (red solid line) initially
    increases in roughly integer steps as the dot is being charged,
    while $n_\downarrow^\QD$ (blue solid line) stays essentially zero,
    because the fixed field $\tB$ is large. However at $\tVcflip =
    0.01303 \tau$ (vertical dashed line) a spin-flip transition
    occurs, where $(n^\QD_\uparrow,n^\QD_\downarrow)$ changes from
    $\simeq (3,0)$ to $\simeq (2,1)$, and the conductance shows a
    spin-flip resonance of height $1$.  (b) Zoom-in of the same data
    to the vicinity of the spin-flip transition at
    $\tVcflip$.}
\end{figure}
For large magnetic fields (beyond about $\tB \gtrsim 10^{-4}$)
  the movie shows several sharp conductance resonances or peaks of
  height $g \simeq 1$, which move in the direction of decreasing
  $\tVc$ (toward the right) with increasing magnetic field.  An
  example of such a resonance, occuring for $\tB = 1.4 \cdot 10^{-3}
  \tau$ and $\tVs = 0.018 \tau$ at $\tVc = 0.01303 \tau$, is shown in
  \Fig{fig:s1s3}.  We will call these ``spin-flip resonances'', since
  their origin lies in spin-flip transitions on the QD; in fact, they
  can be viewed as generalized versions of the singlet-triplet Kondo
  effect discussed in the literature (see \Ref{Pustilnik2003}, and
references therein).  Although the spin-flip resonances
have no relevance for the 0.7-anomaly, they are interesting
in their own right, hence we now explain their origin in some more detail.

With increasing magnetic field, the total spin of a Coulomb-blockaded
QD will increase in discrete steps. This has been discussed in the
  past in terms of the Fock-Darwin spectrum of a QD, see \eg\ Fig.\ 5
  in Ref.\ \onlinecite{Kouwenhoven2001}.  Such a step involves adding
a spin-up electron to the lowest-lying empty dot level while removing
a spin-down electron from the highest-lying doubly-occupied one, which
occurs whenever the gain in Zeeman energy outweights the cost in
kinetic energy. The latter depends on the QD's level spacing, and
hence on $\tVc$ and $\tVs$.  For given $\tB$ and $\tVs$, such a
transition can thus also be induced changing $\tVc$.  To be specific,
decreasing $\tVc$ (as in \Fig{fig:s1s3}) increases the level spacing
and causes a spin-decreasing spin-flip transition, say from the dot
configuration $(n^\QD_\uparrow,n^\QD_\downarrow)$ to
$(n^\QD_\uparrow-1,n^\QD_\downarrow+1)$.  Precisely \textit{at} the
spin-flip transition, say for $\tVc = \tVcflip$, these two
configurations are energetically degenerate, so that Kondo-like
correlations between the QD and the leads can develop, which cause the
conductance $g$ to reach its maximum possible value, namely $1$.

In the movie the heights of these spin-flip resonances typically do
not reach unity, but rather fluctuate as a function of magnetic
field. This is a numerical artefact caused by the insufficient
resolution of $\tVc$ used when making the movie. The numerical effort
that would have been needed to resolve these type of resonances in the
movie would have been very high, since they are typically very narrow.
(We note also that at finite temperature, the minimum width of
these resonances would be set by temperature.)

The fact that $g=1$ at a spin-flip resonance can be understood,
  following \Ref{Pustilnik2003}, using elementary concepts from the
  Fermi-liquid description of zero-temperature transport through a
  multi-level quantum dot.  (For present purposes, we call the entire
  CCR a ``QD'').  Such a description is formulated in terms of the
  eigenphases, say $\delta_{\sigma,1}$ and $\delta_{\sigma,2}$, of the
  scattering matrix of the QD. (A detailed discussion of how these
  eigenphases can be calculated using fRG is given in our follow-up
  paper\cite{Delft2014}.)  These eigenphases are defined \wrt\ a
  reference gate voltage $\tVcr$, at which we set $\delta_{\sigma,1} =
  \delta_{\sigma,2}=0$.  The even and odd linear combinations of these
  eigenphases,
\begin{equation}
\label{eq:define-sigma-plus-minus}
\delta_{\sigma,\pm} = \delta_{\sigma,1} \pm \delta_{\sigma,2} \; ,
\end{equation}
determine, respectively, the quantum dot's charge $n^\QD$ and conductance
$g$.  The charge $n^\QD$, measured \wrt\ the
reference point, is given by Friedel's sum rule:
\begin{eqnarray}
  n^\QD
  & = & \hspace{-1mm} \phantom{\frac{1}{2}} \;
  n^\QD_\uparrow + n^\QD_\downarrow \;
  = \; \frac{1}{\pi} \; \left( \delta_{\uparrow,+} + \delta_{\downarrow,+} \right) .
\end{eqnarray}
The conductance is given by the relation
\begin{eqnarray}
\label{eq:FL-conductance}
g  & = & \frac{1}{2} \sum_\sigma \sin^2 \delta_{\sigma,-} \; ,
\end{eqnarray}
valid for left-right symmetric couplings between QD and leads,
as in our model. We
choose $\tVcr$ to lie near the spin-flip transition, but sufficiently
different from $\tVcflip$ that the conductance and the phases
determining it are small at $\tVcr$, i.\,e.\ $g \ll 1$ and
$|\delta_{\sigma, -}| \ll \pi/2 $.  

Now, when $\tVc$ is tuned through the spin-flip transition, the
$\tVc$-induced changes in $\delta_{\sigma,-}$ and $n^\QD_\sigma$
(\wrt\ to their values, namely 0, at the reference voltage $\tVcr$)
are related by
\begin{eqnarray}
\label{eq:changes-in-delta-minus}
\Delta \delta_{\sigma,-} \simeq \pi \Delta n^{\QD}_\sigma \; .
\end{eqnarray}
This equation follows from two facts: first, one of the spin-dependent
eigenphases of the scattering matrix, either $\delta_{\sigma, 1}$ or
 $\delta_{\sigma, 2}$, turns
out to be essentially independent of $\tVc$ throughout a
Coulomb-blockade valley, so that
$|\Delta \delta_{\sigma,-} |= |\Delta
\delta_{\sigma,+}|$ [by \Eq{eq:define-sigma-plus-minus}]; second, the
Friedel sum rule implies that $\Delta \delta_{\sigma,+}= \pi \Delta
n^\QD_\sigma$.  Now, since the total dot charge $n^\QD =
n^\QD_\uparrow + n^\QD_\downarrow$ is fixed within the Coulomb
blockade valley, the spin-dependent dot occupancies change in equal
but opposite manner as $\tVc$ is tuned through the spin-flip
transition: $\Delta n^\QD_\uparrow \simeq - \Delta
n^\QD_\downarrow$. By \Eq{eq:changes-in-delta-minus}, this implies
that both $|\Delta \delta_{\uparrow, -}|$ and $|\Delta \delta_{\downarrow,
  -}|$ will pass through $\pi/2$ at essentially the same value of
$\tVc$, causing the conductance $g$ [\Eq{eq:FL-conductance}] to show a
resonance of height $\simeq 1$ there.
%(If $n^\CCR$ is not precisely an integer,
%then $n^\CCR_\uparrow$ and $ n^\CCR_\downarrow$ will cross through
%half-integer values at slightly different values of $\tVc$, and the
%height of the conductance resonce will be somewhat reduced below
%$\GQ$.)
The case shown \Fig{fig:s1s3} is an example of a so-called
``triplet-singlet'' transition\cite{Pustilnik2003}, where the spin of
the QD changes from 3/2 to 1/2 as $\tVc$ decreases past $\tVcr$. 

%\bibliography{../kondo07}

\end{document}